\documentclass[twocolumn,floatfix,showkeys,preprintnumbers,amsmath,amssymb]{revtex4}

\usepackage{graphicx}
\usepackage{subfigure}
\usepackage{dcolumn}
\usepackage{bm}

\begin{document}

\title{Status and Scope of MONC Transport Code\\}
\author{H. Kumawat$^1$\footnote{author. Email address: harphool@barc.gov.in}}
\affiliation{$^1$Nuclear Physics Division, BARC, Mumbai-400085, India\\}
\author{P.P.K. Venkata$^2$}%
\affiliation{$^2$Computer Division, BARC, Mumbai-400085, India\\}
\date{\today}

\begin{abstract}
$\underline{\textbf{MO}}$nte-carlo $\underline{\textbf{N}}$ucleon transport $\underline{\textbf{C}}$ode (MONC) for nucleon transport is being developed for several years. Constructive Solid Geometry concept is applied with the help of solid bodies. Union, subtraction and intersection Boolean operations are used to construct heterogeneous zones. Scaling, rotation, and translation operation of the basic bodies are allowed to construct more complex zones. Module of repetitive structure for lattice, core calculations in reactor and detector simulation is developed. Graphical User Interface along with visualization tools is developed to make input, construction and display of geometry, and analysis of output data. Low energy neutron transport module is developed using continuous linearly interpolable point neutron cross section data below 20MeV neutron energy. The code is benchmarked for simulation of accelerator driven sub-critical system, neutron shielding, heat and neutron flux distribution and keff of the critical assemblies. It is observed that results of keff  are in agreement within $\sim$ 3mk with experimental results of critical assemblies as well as the values obtained from MCNP.
\end{abstract}

\maketitle
\section{\label{sec:level1}Introduction}
In past two decades development in the accelerator technology has given boost to facilitate the world with Spallation
Neutron Source (SNS) and to study Accelerator Driven Sub-critical systems (ADS) \cite{ads1,ads2,ads3}. The essential component of an ADS system is the presence of a spallation target in the core of the reactor. 
An external proton beam (energy ~ 1 GeV and current ~ tens of mA) produces spallation neutrons 
which drives the reactor under sub-critical conditions. Currently two types of generic target modules i.e. 
window and windowless concepts have been proposed \cite{twindow1,twindow2,twindow3}. In the window configuration, the proton beam enters the target region through a thin solid barrier (window), which isolates the beam transport pipe and the target. The spallation target is ideally conceived to be material of high atomic number (Z) or higher (N/Z) ratio and low melting point like a lead bismuth eutectic (LBE) while a few mm thick steel is used as window material. One of the critical operating conditions of the spallation target is the very high volumetric heat deposition rate. For example, ~65$\%$ proton beam energy at 1 GeV is lost as ionization energy which is just heating the spallation target system. 
Depending on the density, the rate of heat deposition is also different for both target and window materials. The heat loss at a given energy is sum of two components; the contributions coming from direct ionization of primary beam and ionization through the hadronic interactions. At low energy, as the hadronic contribution is not very significant, the total heat loss is predominantly due to primary ionization.  The longitudinal heat loss profile (dE/dz) as a 
function of z has a range rb followed by a Bragg peak, an interaction process which can be described well by the 
Bethe formula. However, at 1 GeV proton energy, the primary ionization accounts for only 70$\%$ of the total heat 
loss and the remaining 30$\%$ comes due to hadronic interactions (High energy fission, charged evaporation products,
$\pi^0$ decay, nuclei de-excitation, slowed down charged particles, nuclei recoil). This hadronic contribution goes 
up with incident beam energy and the Bragg peak is no longer visible.  At 1 GeV, the dE/dz shows a maximum in the 
spallation zone very close to the window material and decreases thereafter with increasing z. Since the dE/dz is
maximum in the window region, it is exposed to the highest temperature, hence it is important to study heat loss 
in the window very carefully to optimize flow and geometry parameters for effective cooling of the beam window.  
In an earlier work \cite{hkheat08},  we had carried out thermal hydraulic studies using computational fluid dynamics (CFD) 
code related to the design of a realistic LBE target for a one way coupled 750 MW  thermal reactor which requires 
current in the range of 1-2 mA proton beam of 1 GeV energy. The input to the CFD code; the heat deposit profile was estimated using CASCADE.04 \cite{hkumawat04, hkumawat05} and FLUKA Monte-Carlo simulation codes. It was realized that the precise knowledge of energy 
loss dE/dz  in the window material is very crucial for the target design as any small variation can result
in a different set of operating parameters. Therefore, in the above mentioned publication, we gave more emphasis on the study of 
energy deposition due to interaction of high energy proton in various thick targets and compared the results 
with available experimental measurements. As a bench mark study, the results obtained from both CASCADE.04 and FLUKA are compared with dE/dz measurements for Be, Al, Fe, Cu, Pb and Bi thick targets of proton energies 0.8 GeV, 1.0 GeV and 1.2 GeV. 
It was found that both FLUKA and CASCADE.04 results were comparable with the experimental measurements 
for heavy targets like Pb and Bi where as for targets like Be and Al, both the codes as well 
as experimental measurements show differences.

The CASCADE.04 code was further developed for low energy neutron transport with point data processed through NJOY (ACE format). The code was also developed for the complex geometry using boolean operation on Constructive Solid Geometrical Shapes and for repetitive reactor core/detector or any other complex structures. Reactor or ADS neutron multiplication factor/K$_{eff}$/K$_{\infty}$ capabilities and GUI were developed \cite{hkvenkata13}. We have also developed decay module using linear chain method. With all these development the code was released for public use through a DAE-BRNS Workshop on MOnte Carlo Nucleon transport Code (MONC) BARC, Oct. 8-9, 2015 held at BARC Mumbai. We have also developed our own processing code in ROOT format files \cite{hkroot} under BARC-CERN collaboration for GeantV. The Decay code is extended to do reactor burnup calculations. The average cross-section for burnup is calculated during monte-carlo calculation after burnup cycle and updated for the next cycle while updating the inventory of fission fragments. 

The outline of this paper is as follows. In Sec. II we present description of Geometry. Section III contains High energy module of the code. In
section IV Low energy transport module is presented. In Sec. V, the
method of criticality calculations are given. In Sec. VI
Criticality benchmark calculations are presented. In Sec. VII
Decay and burnup calculations are presented. Section IX considers Graphical user Interface. Conclusions are
given in Sec. X.

\section{\label{sec:level2}Construction of Geometry}
\begin{figure}
\includegraphics[scale=0.8]{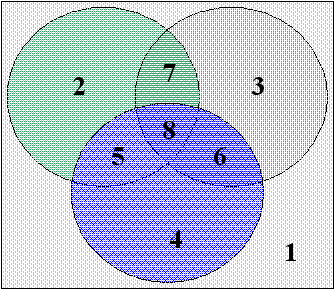}
\caption{CSG model depicting different zones during geometry construction in MONC} \label{csg1}
\end{figure}

Most important and difficult task in the Monte Carlo code is to build a complicated geometry 
in easiest and user friendly manner. We have chosen CSG model to build the geometry. 
In this model, there are simple basic geometrical bodies viz. Sphere, Cylinder, 
Parallel-piped, Cone, Ellipse, Hexagon etc. Boolean operations (Union, subtraction 
and interaction) are enabled to construct complex zones using these bodies. 
Fig. \ref{csg1} gives an example to construct the zones from the bodies. One universe is 
compulsory to define that bounds the whole geometrical structures. 
Eight heterogeneous zones from three spherical bodies and one universe are made which can be filled with different materials. 

\begin{figure}
\includegraphics[scale=0.8]{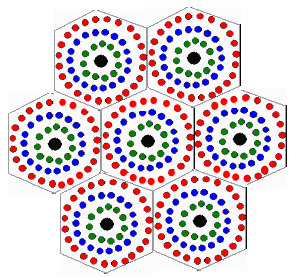}
\caption{Repetitive CSG model during geometry construction in MONC} \label{csg2}
\end{figure}

Scaling, rotation and translation of the bodies are used in the chronological 
form to make more complicated structures. Repeated geometry structure has been 
invoked to do the lattice and core calculations of the most complicated reactor
assemblies. Fig. \ref{csg2} shows an example of repeated structure which are filled from 
different type of fuel rods. Hexagonal bounding boxes can be logical or real. 
Bare minimum information viz. number of rods, mean radius of ring on that rods 
are to be placed,  and center of the repeated structures has to be provided by 
the user. Reflective boundary conditions are used to do single lattice or partial core calculations.

\section{\label{sec:level3} High energy particle transport}
Monte Carlo program MONC incorporates Intra-nuclear Cascade, Pre-equilibrium, 
Evaporation and Fission models to simulate spallation reaction mechanism for 
thin and thick targets. Modeling details of Intra-nuclear cascade, Pre-equilibrium 
particle emission are described in detail in Ref. \cite{baras1, baras2}. Treatment of cutoff energy 
from Intra-nuclear to pre-equilibrium and next to evaporation stage was changed 
later which is described in Ref. \cite{hkumawat05}. Generalized evaporation model was developed 
as described in Ref. \cite{mashnik} and Fong’s statistical fission model is used to simulate 
the high energy fission reaction. Fission barrier, level density parameter and 
inverse cross sections for pre-equilibrium/evaporation/fission model are given in detail in Ref. \cite{hkumawat04, hkumawat05}.

\begin{figure}
\includegraphics[scale=0.4]{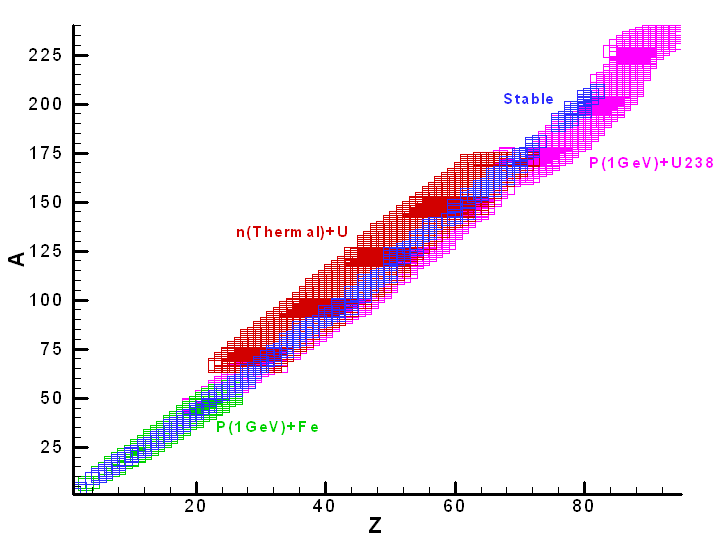}
\caption{Proton induced spallation and neutron induced fission product yields calculated using MONC.
Stable isotopes are marked with blue, p+$^{238}U$ spallation products are marked with pink, fission products for n(thermal)+$^{235}U$ are shown by red color and green color shows spallation products from p+$^{56}Fe$.} \label{spalfisprod}
\end{figure}

Benchmark of spallation models for experimental values of neutron, charged particles, 
and pions double differential production cross-sections, particle multiplicities, 
spallation residues and excitation functions was organized by IAEA and is given in Ref. \cite{hkumawat10}. 
We have used the predecessor of this code named CASCADE.04 to calculate these quantities in 
this benchmark. Heat Deposition algorithm for thick spallation targets and thin films was
modified and benchmarked as mentioned in Ref. \cite{hkheat08}. The code was further developed for 
the Neutron shielding and dosimetry applications and published \cite{hkumawat09}. The high energy part 
of this code can be used for single nucleus interaction in basic reaction studies and 
can be invoked for the thick target simulation during the transport. Energy loss of 
the charge particle is calculated during the transport in the thick target. 
The code calculates the spallation as well 
as fission yields which are plotted in Fig. \ref{spalfisprod}. It is clear from Fig.4 that spallation 
products are proton rich but neutron induced fission products are neutron rich. 
The beauty of the spallation reaction is that it can span whole range of radio active 
isotopes around the stability line. Spatial distribution of any toxic/nontoxic element 
can be analysed and is available as an option in the input file.

Monte Carlo (MC) program CASCADE-04 (Intra-nuclear cascade – Preequilibrium - Evaporation/ Fission code) 
realizes the particle transport in three stages: 
1) sampling of particle (ion) mean free path in the medium taking into account the energy loss of a 
charged particle and a possible decay of non-stable particles ($\pi^0$, $\pi^\pm$). All $\pi^0$-mesons are considered to decay into $\gamma$-quanta at the point of their creation. The ionization losses of $\pi$ - mesons, protons and light ions are calculated by Sternheimer’s method \cite{stern} using well established Bathe formula for the average ionization loss calculations with proper density effects. Here, it is important to mention that the density effect shows reduction in ionization loss for fast charged particles due to dielectric polarization of the medium. In the lower energy region (< 2.0MeV) Lindhard’s approach \cite{lind} is used and a semi-phenomenological procedure \cite{barasion} is applied for the heavy ions. 
While doing the practical simulation one has to calculate the ionization and nuclear interaction ranges 
and then uses the formulation to deposit heat. It is well known that heat deposition is a continuous 
phenomenon but in simulation one has to take some finite step size. In the earlier version (CASCADE.04) 
of the code we have set no limit to heat deposition step size and it may deposit total amount of heat 
at the ionization range point which is impractical. Of course it was provision that instead of depositing 
the heat at ionization range point it will deposit at the middle of last two interaction points. 
In this situation we found that the heat inside few centimeter depth of the target was less than 
the experimental data. In CASCADE.04.h we have adopted more continuous steps in which not more than 
2-3\% energy loss can appear at any point. In this way we could get better agreement with 
experimental data.

2) Simulation of the particle interaction with a nucleus is considered along its path. 
In case of inelastic interaction the CASCADE.04 code considers three stages of reaction for calculation: 
a) intranuclear cascade originally developed at Dubna: In this part of the calculation,
primary particles can be re-scattered and they may produce secondary particles several
times prior to absorption or escape from the target. Modeling of intra-nuclear cascades 
\cite{baras1, baras2} is in general rather closer to the methods used in other transport codes.
Cross-sections of the hadron-nucleus collisions are calculated based on the compilations 
of the experimental data \cite{barascrs1, barascrs2}. To calculate the nucleus-nucleus cross-sections 
we used analytical approximations with parameters defined in ref. \cite{barascrs3}. Criteria of 
transition from intra-nuclear cascade to pre-equilibrium stage are the cutoff energy 
(binding energy above the Fermi energy), below which the particles are considered to
be absorbed by the nucleus. Particles are traced down to this cutoff energy and then 
the second stage, pre-equilibrium starts, b) Pre-equilibrium stage: In this part of 
the reaction, relaxation of the nuclear excitation is treated according to the 
exciton model of the pre-equilibrium decay. The relaxation is calculated by the 
method based on the Blann's model\cite{mashnik, blann}. Proton, neutron, deuterium, tritium, 
3He and 4He are considered as emitted particles in the pre-equilibrium and in the
subsequent equilibrium stage. Transition from pre-equilibrium to equilibrium state
of the reaction occurs when the probability of nuclear transitions changing the 
number of excitons n with $\delta$n=+2 becomes equal to the probability of transitions 
in the opposite direction, with $\delta$n=-2, c) Equilibrium stage: This part considers 
the particle evaporation/fission of the thermally equilibrated nucleus.


\section{\label{sec:level5} Low energy neutron transport}
Low energy neutron transport code is developed recently. We have developed the package for reading pointwise cross sections for neutron in ACE (A Compact ENDF) format. The delayed neutrons are treated exclusively with their energy spectra for which data are available. Spontaneous and induced fission fragment yield are read from ENDF Fission yield libraries. The ACE library generated, mostly, using ENDF VII.0 is used for the present investigations. The free gas thermal treatment of the
neutron interaction for below 4eV can be used for compound and crystal material or Thermal scattering law can be used if available in ENDF file. Probability table method is used in the un-resolved energy region.

Interaction of neutrons is considered using Monte Carlo method in the following steps.
A) Identification of the initial zone number and interaction cite of the neutron,
B) Selection of the collision nuclide,
C) Type of interaction. 

\subsection{Identification of the initial zone number and
interaction cite of the neutron}

The MONC assigns the X, Y, Z, cos$\theta$, sin$\phi$, cos$\phi$,
energy(MeV), charge, mass(MeV/C$^2$) coordinates with each
neutron. The code identifies the zone number and geometry type of
the configuration defined in the input file by calculating the
surface coefficients. The method of neutron transport is as given
below. The probability of interaction between l and l+dl of a
neutron of given energy is defined by

\begin{equation}
p(l)dl=exp^{-\Sigma_tl}\Sigma_t dl \label{lowp0}
\end{equation}

Where $\Sigma_t$ is the macroscopic total cross-section of the
medium in the geometrical Zone. Mean free path for the nuclide inside the the medium of a
given composition can be given as.

\begin{equation}
mean free path = \frac {1.}{\Sigma_t } ; \Sigma_t =
\frac{1.}{n\sigma_t } \label{lowp1} \end{equation}

Where n is the nucleon density of the selected nuclide. If  $\xi$
is random number between [0-1] then

\begin{equation}
 \xi = \int
\limits_{0}^{l}exp^{-\Sigma_ts}\Sigma_t ds =
1.-exp^{-\Sigma_tl}\label{lowp2} \end{equation}

 Thus \begin{equation}
 l = -\frac
{1}{\Sigma_t}ln(1-\xi ) \equiv -\frac {1}{\Sigma_t}ln(\xi )
\label{lowp3} \end{equation}

\subsection{Selection of the collision nuclide}

If there are n different nuclides forming the material composition
in the identified zone then i$^{th}$ nuclide is selected if
\begin{equation}
\sum\limits_{j=1}^{i-1}\Sigma_{tj} <\xi  \sum \limits_{j=1}^{n-1}
\Sigma_{tj}\leq \sum \limits_{j=1}^{i}\Sigma_{tj} \label{lowp4}
\end{equation}

Where $\Sigma_{tj}$ is the macroscopic total cross-section of
the j$^{th}$ nuclide. The MONC code needs to describe the density
(g/cm$^3$) and nuclei weight fractions in the input file. The sum
of all fractions should be normalized to one bu user otherwise code will re-normalize
it to one.

\subsection{Type of the interaction}
Type (elastic, inelastic, capture, fission) of interaction with
the selected nuclide is calculated directly based on the
probabilities, calculated using the microscopic cross-sections.
These reaction probabilities can be given as eq. \ref{lowp5}

\begin{equation}
 \frac {\sigma_{i}}{\sigma_t }\label{lowp5}
\end{equation}

Where subscript 'i' stands for reaction type (elastic, inelastic,
capture, fission, (n,xn), (n,X)) etc. and 't' for total cross-section.
In case of isotropic angular distribution of elastic scattered
neutrons in the center of mass system, the energy has been
calculated using two body collision kinematics and then converted
from CM to LAB system.

\section{MONC Criticality Calculations}

The critical calculations in MONC are based on four methods (neutron population, 
Collision Estimator, Absorption Estimator, and Track Length Estimator). 
The k$_{eff}$ is a ratio between the numbers of neutrons in successive generations in a 
fission chain reaction. For critical systems, k$_{eff}$ = 1, for sub-critical systems, 
k$_{eff} <$ 1 and for supercritical systems, k$_{eff} >$ 1. The number of neutrons in successive 
generation is obtained from number of neutrons generated by fission minus the number 
of neutrons absorbed and escaped from the system. Whenever (n, xn) reactions occur, 
the neutron generated are again transported within the fission cycle. At present 
fission source points as well as neutron generations are as usual allowed as other 
reactions but stored for the next cycle. At the end of the cycle same number of 
neutrons are preserved by increasing or decreasing the weight of neutrons in case of (k$_{eff} >$ 1) and (k$_{eff} <$ 1), respectively.

The critical calculation requires number of inactive cycles
which need to be skipped to get the fundamental mode of fission source, active 
cycles for actual k$_{eff}$, and number of source neutrons. Mono energetic neutron can 
be defined very easily in the input file and spectrum can be given through a
separate file. In case of high energy proton or other beam, the source distribution 
is generated using the high energy part of the code and that is transported below 20 MeV.
The value of k$_{eff}$ is based on these low energy neutron fission reactions. 
The average value of k$_{eff}$ is obtained from the maximum likelihood method applied 
for the values obtained from all four estimators. 

\subsection{neutron population method}
The ratio of previous to next fission cycle while considering absorption and escape, gives value of 
k$_{eff}$ as given below.
\begin{equation}
 k_{eff}=\frac {\textmd{neutron balance in the i+1}^{th} \textmd{generation}}
 {\textmd{neutron balance in the i}^{th} \textmd{generation}}\label{crit1}
\end{equation}
\subsection{Collision Estimator}
It is based on the collision/fission reaction in the fission zones.
\begin{equation}
 k_{eff}=\frac{1}{N}\sum_{i}W_{i}\frac {\sum_{k}f_{k}\nu_{k}\sigma_{fk}}
 {\sum_{k}f_{k}\sigma_{Tk}}\label{crit2}
\end{equation}
where i is summed over all collisions in a cycle if fission is
possible, k is summed over all nuclides of the material involved
in the i$^{th}$ collision,$\sigma_{Tk}$ = total microscopic cross
section, $\sigma_{fk}$= microscopic fission cross section,
$\nu_{k}$= average number of prompt or total neutrons produced per
fission by the collision nuclide at the incident energy, f$_k$ =
atomic fraction for nuclide k, N = source neutrons for cycle, and
W$_i$ = weight of particle entering collision.

\subsection{Absorption Estimator}
It is based on the absorption in fission and other reactions in the fission zones.
\begin{equation}
 k_{eff}=\frac{1}{N}\sum_{i}W_{i}\nu_{k}\frac {\sigma_{fk}}
 {\sigma_{ck}+\sigma_{fk}}\label{crit3}
\end{equation}
where i is summed over all fissions, k is summed over all nuclides
of the material involved in the i$^{th}$ collision,$\sigma_{ck}$ =
capture cross section, $\sigma_{fk}$= microscopic fission cross
section, $\nu_{k}$= average number of prompt or total neutrons
produced per fission by the collision nuclide at the incident
energy, N = source neutrons for cycle, and W$_i$ = weight of
particle entering collision.

\subsection{Track Length Estimator}
This method uses length of the neutron track in the fission zones.
\begin{equation}
 k_{eff}=\frac{1}{N}\sum_{i}W_{i}\rho d \sum_{k}f_{k}\nu_{k}\sigma_{fk}\label{crit4}
\end{equation}
where i is summed over all trajectories in a cycle where fission
is possible, k is summed over all nuclides of the material
involved in the i$^{th}$ collision, $\rho$ = aromic density in the
region, d= track length.
\subsection{Combined k$_{eff}$ Estimation}
The average value of k$_{eff}$ is estimated using the Maximum
Likelihood method from all four estimators. It is recommended that
the user should run for at least two histories (having two active and inactive cycles) to get better standard deviation. The user should also watch that the k$_{eff}$ values by
all methods should not very more than 0.1\%, also that the values
should not wonder too much from the average values otherwise the
number of source particles or the number inactive cycles has to be
increased.
\section{Benchmark Calculations}

\subsection{IAEA-ADS Benchmark Problem}
\begin{figure}
\mbox{\subfigure{\includegraphics[scale=0.25]{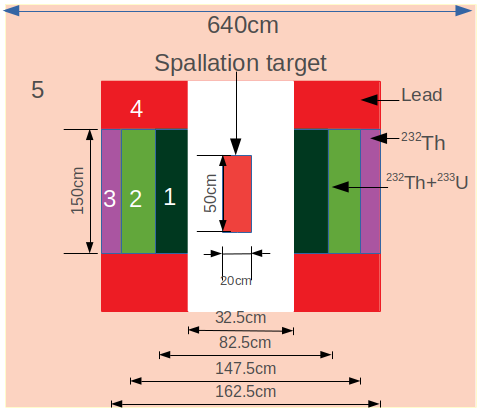}}}
\quad\subfigure{\includegraphics[scale=0.25]{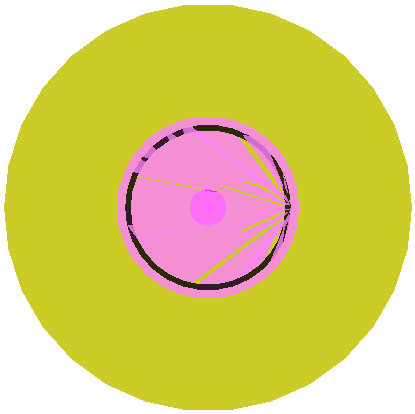}}
\caption{{side and Front view of the IAEA-ADS benchmark assembly}} \label{adsbench1}
\end{figure}

IAEA had organized a coordinated research project for international neutronics benchmark of 
an ADS with $^{233}$U-$^{232}$Th Fuel. The system is driven by an external spallation neutron source.
A proton beam of energy 1.0GeV hits Lead as spallation target placed in the center of the ADS.
The spatial and energy distribution of the spallation neutrons are given in the benchmark
but we have produced these neutrons from our code itself. Geometry of ADS is divided into 5 regions
as shown in FIG.\ref{adsbench1}. The nuclei density at BOL is gvien in Table \ref{tab:nucleiads}.
The temperature of fuel was taken as 1200K and that of Lead and steel was 900K. 
\begin{table}
\caption{\label{tab:nucleiads} Nuclei densities (/barn-cm) in different regions of the ADS benchmark at BOL}
\begin{ruledtabular}
\begin{tabular}{ccccccc}
Nuclei              &  region 1  & region 2   &region 3  &region 4   &region 5\\\hline
$^{232}$Th          &            &            & 7.45E-03 &           &        \\
$^{233}$U+$^{232}$Th& 6.35E-03   & 7.45E-03   &          &           &        \\
O                   & 1.27E-02   & 1.49E-02   & 1.49E-02 &           &        \\
Fe                  & 8.10E-03   & 8.87E-03   & 8.87E-03 &           &  6.63E-03      \\
Cr                  & 1.12E-03   & 1.06E-03   & 1.06E-03 &           &  8.00E-04      \\
Mn                  & 4.60E-05   & 5.10E-05   & 5.10E-05 &           &  3.80E-05      \\
W                   & 4.60E-05   & 5.10E-05   & 5.10E-05 &           &  3.80E-05      \\
Pb                  & 1.77E-02   & 1.56E-02   & 1.56E-02 & 3.05E-02  &  2.41E-02      \\
\end{tabular}
\end{ruledtabular}
\end{table}

The first thing is to find the $^{233}$U enrichment for k$_{eff}$ = 0.98, 0.96 and 0.94 at BOL.
The enrichment \% are given in Table \ref{tab:u233enrich}.
\begin{table}
\caption{\label{tab:u233enrich} $^{233}$U enrichment at BOL for ENDF-B/VIII.1, ENDF-B/VII.1 in the region 1 and 2.}
\begin{ruledtabular}
\begin{tabular}{ccccccc}
Participants            &  k$_{eff}$ = 0.98  & k$_{eff}$ = 0.96   & k$_{eff}$ = 0.94\\\hline
MONC-2.0                & 10.04   & 9.65   &   9.38          \\
Russia (diffusion)      & 10.01   & 9.69   &   9.38  \\
Russia (MC)             & 10.26   & 9.925  &   9.61  \\
Switzerland             & 9.88    & 9.57   &   9.25  \\
Italy                   & 10.29   & 9.96   &   9.63  \\
France                  & 10.27   & 9.94   &   9.61  \\
German                  & 10.00   & 9.68   &   9.36  \\
Netherlands             & 10.13   & 9.81   &   9.49  \\
Japan                   & 9.7     & 9.4    &   9.1  \\
Belarus                 &  10.50  & 10.17  &   9.85    \\
Sweden                  &  10.419 & 10.095 &   9.771   \\
Average                 &  10.17  & 9.85   &   9.53    \\
MCNT/ORIGEN2            &  9.99   & 9.672  &   9.353   \\
ANDOTOR                 &  10.265 & 9.948  &   9.629  \\
\end{tabular}
\end{ruledtabular}
\end{table}

\begin{figure}
\includegraphics[scale=0.4]{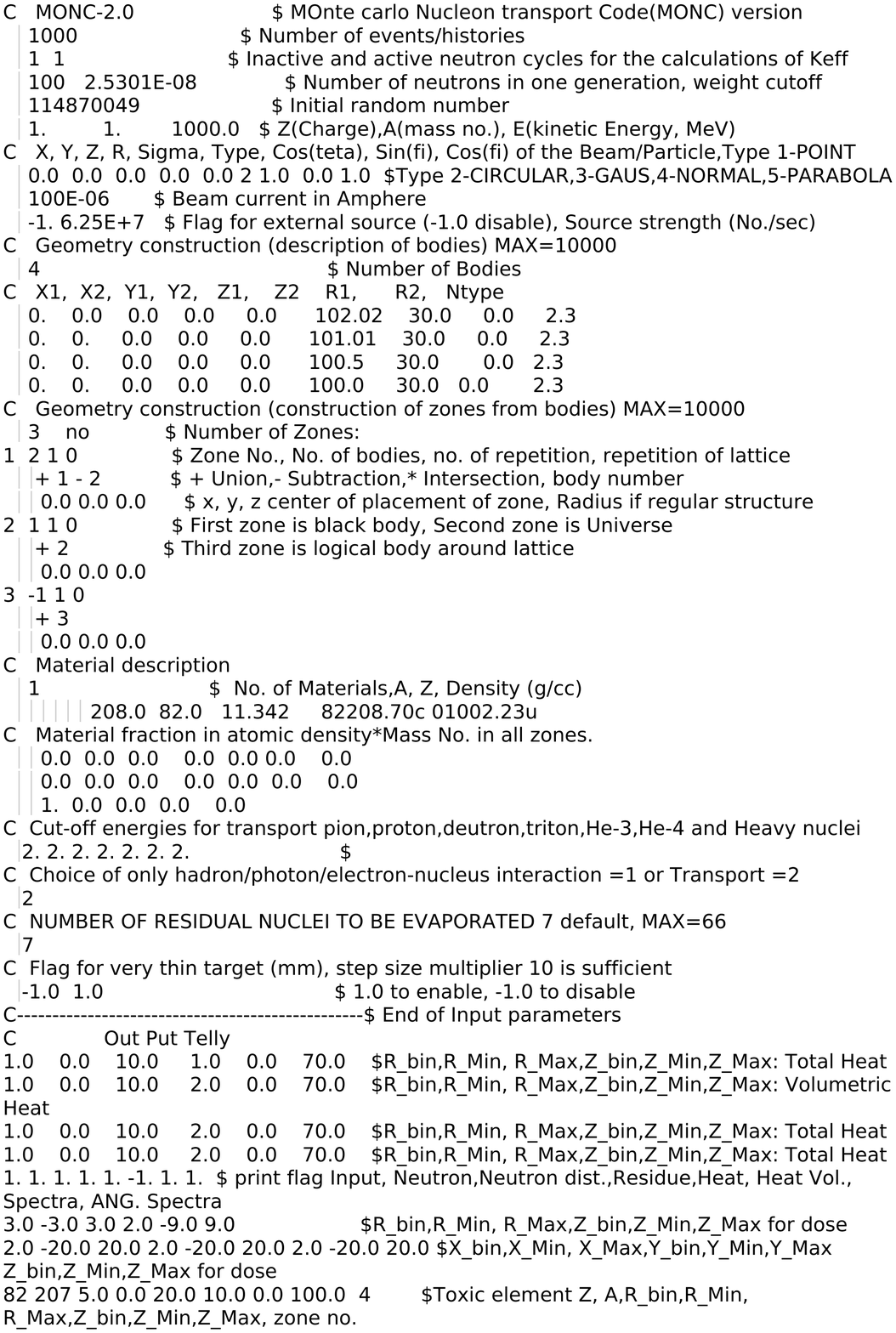}
\caption{Input file for spallation target} \label{inpspal}
\end{figure}

Sample input file with some comments for the high energy spallation reaction is given in FIG. \ref{inpspal}

\subsection{Some sample problem for k$_{eff}$ calculations}
Benchmark criticality problem as defined in ref. \cite{godivarep} are calculated and compared.

1)The Enriched U-235 93.71\% sphere of radius 8.741cm consisting
of 52.42kg mass and density 18.74g/cc. U-238 is 6.29\% by weight.

k$_{eff}$(MONC)=0.9953 $\pm$ 0.0018

k$_{eff}$(MCNP)=0.9962 $\pm$ 0.0009

\begin{figure}
\includegraphics[scale=0.4]{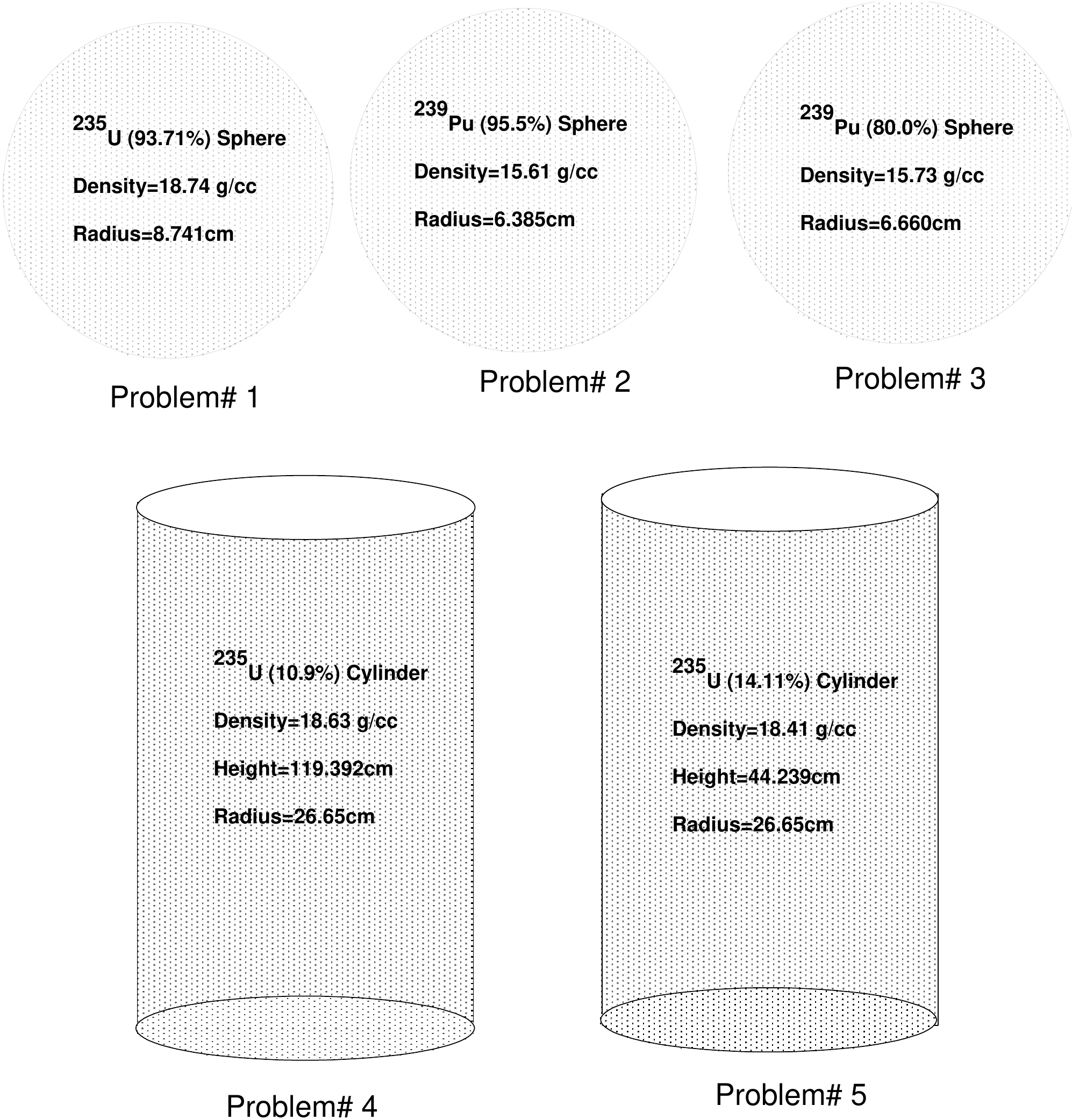}
\caption{Different Spherical and Cylindrical Critical assemblies}
\label{godiva}
\end{figure}

2)The Enriched Pu-239 95.5\% sphere of radius 6.385cm consisting
of 17.02kg mass and density 15.61g/cc. U-240 is 4.5\% by weight.

k$_{eff}$(MONC)=1.0047 $\pm$ 0.0003

k$_{eff}$(MCNP)=1.0052 $\pm$ 0.0006

3)The Enriched Pu-239 80.0\% sphere of radius 6.660cm consisting
of 19.46kg mass and density 15.73g/cc. U-240 is 20.0\% by weight.

k$_{eff}$(MONC)=1.0063 $\pm$ 0.0004

k$_{eff}$(MCNP)=1.0097 $\pm$ 0.0002

4)The Enriched U-235( 10.9\%) Cylinder of radius=26.65cm,
Height=119.392cm and density=18.63g/cc. Remaining material is
U-238.

k$_{eff}$(MONC)=0.9900 $\pm$ 0.0018

k$_{eff}$(MCNP)=0.9915 $\pm$ 0.0005

5)The Enriched U-235( 14.11\%) Cylinder of radius=26.65cm,
Height=44.239cm and density=18.41g/cc. Remaining material is
U-238.The geometry for problems 1-5 are shown in fig.\ref{godiva}

k$_{eff}$(MONC)=0.9905 $\pm$ 0.0018

k$_{eff}$(MCNP)=0.9908 $\pm$ 0.0006

\begin{figure}
\includegraphics[scale=0.45]{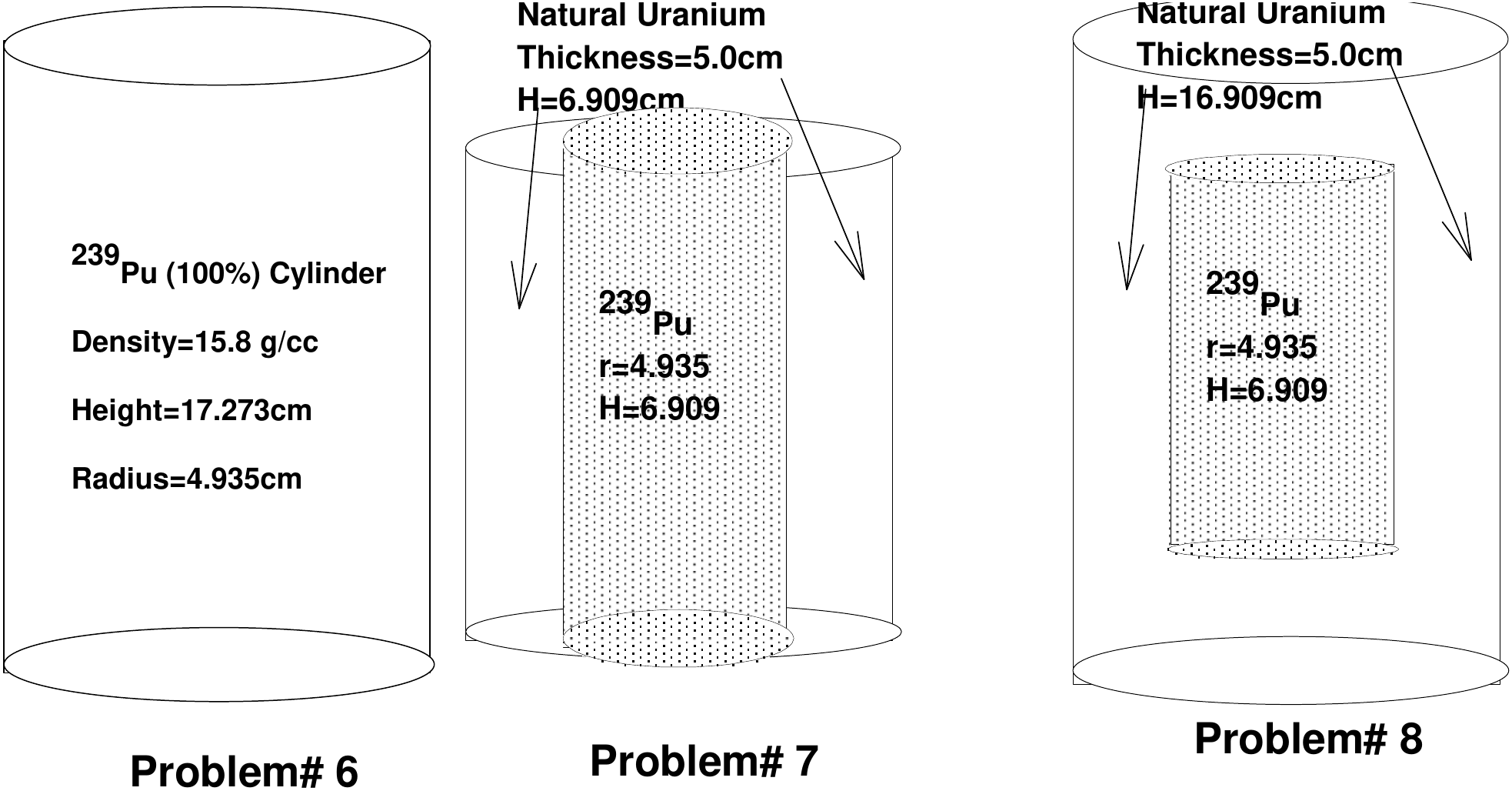}
\caption{Different Cylindrical Critical assemblies}
\label{godiva1}
\end{figure}
6)The Enriched Pu-239( 100.0\%) Cylinder of radius=4.935cm,
Height=17.273cm and density=15.8g/cc.

k$_{eff}$(MONC)=1.0142 $\pm$ 0.0005

k$_{eff}$(MCNP)=1.0157 $\pm$ 0.0003

7)The Enriched Pu-239( 100.0\%) Cylinder of radius=4.935cm,
Height=6.909cm and density=18.80g/cc. Uranium Reflector Thickness=
5cm, Height=6.909cm.

k$_{eff}$(MONC)=0.8879 $\pm$ 0.0018

k$_{eff}$(MCNP)=0.8865 $\pm$ 0.0001

8)The Enriched Pu-239( 100.0\%) Cylinder of radius=4.935cm and
density=15.8g/cc. Uranium Reflector Thickness= 5cm all around
(radially and axially), Density=18.8g/cc. The geometry for
problems 6-8 are shown in fig.\ref{godiva1}

k$_{eff}$(MONC)=1.0225 $\pm$ 0.0004

k$_{eff}$(MCNP)=1.0248 $\pm$ 0.0006

\begin{figure}
\includegraphics[scale=0.4]{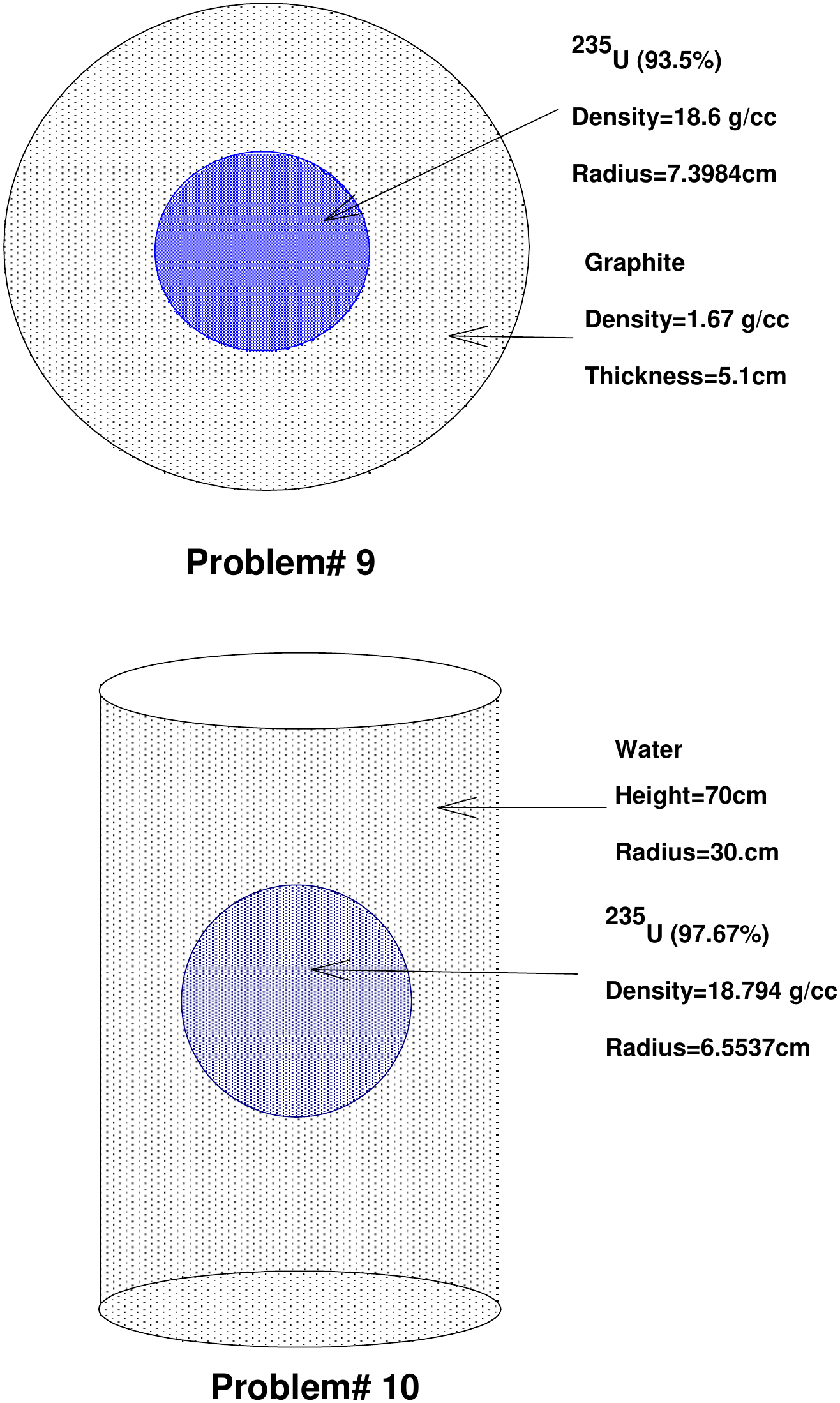}
\caption{Different Critical assemblies} \label{godiva2}
\end{figure}

9)The Enriched U-235 93.5\% sphere of radius 7.3984cm and density
18.6g/cc. The sphere is surrounded by graphite of 5.1cm thickness.
The graphite consists of 99.5\% Carbon, 0.34\% iron and 0.16\%
sulfur. Density of graphite is 1.67g/cc.

k$_{eff}$(MONC)= 0.9983$\pm$0.00010

k$_{eff}$(MCNP)=0.9981 $\pm$ 0.0029

10)The Enriched U-235 97.67\% sphere of radius 6.5537cm consisting
of 22.16kg mass and density 18.794g/cc. The sphere is surrounded
by water tank of radius 30cm and height 70cm. The geometry for
problems 9-10 are shown in fig.\ref{godiva2}

k$_{eff}$(MONC)= 0.9981$\pm$ 0.007

k$_{eff}$(MCNP)=0.9956 $\pm$ 0.0011

\begin{figure}
\includegraphics[scale=0.45]{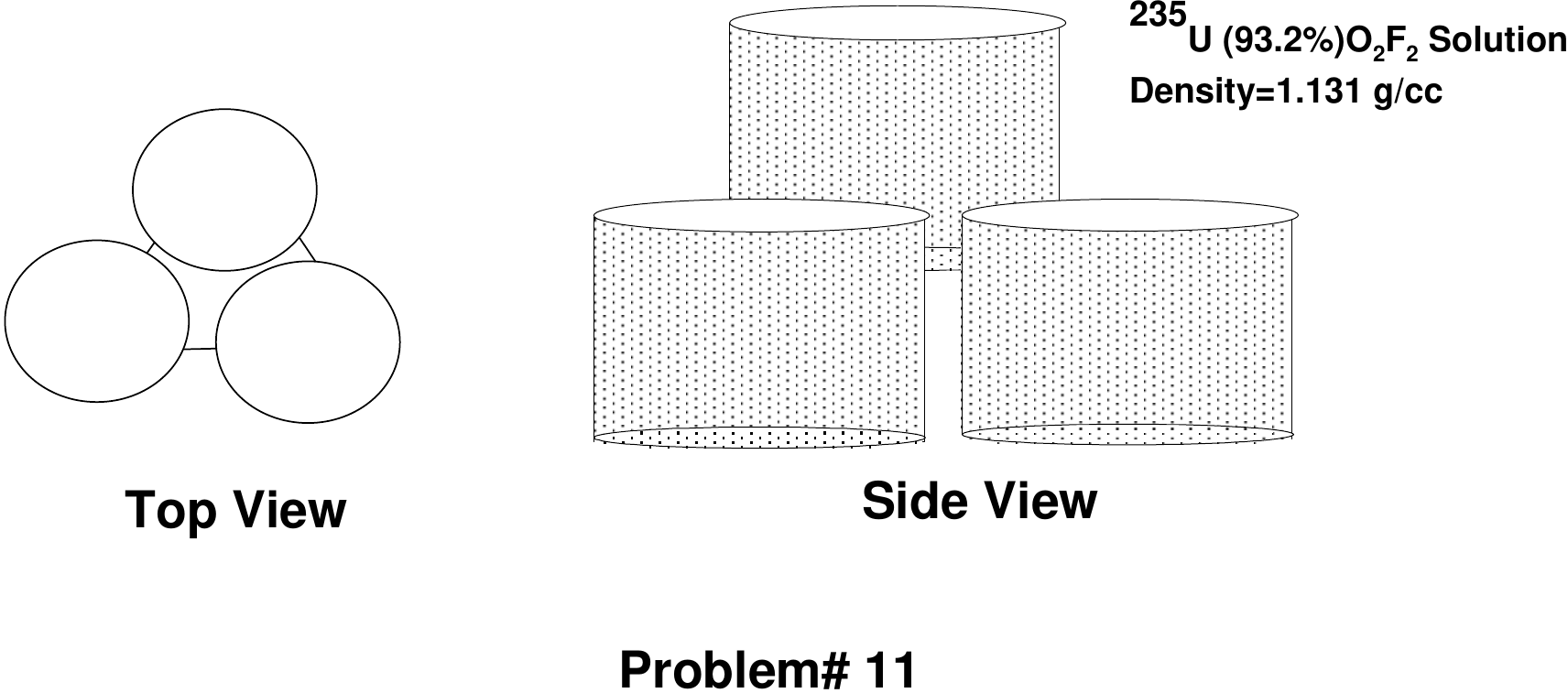}
\caption{Different Critical assemblies} \label{godiva3}
\end{figure}

11)Three interacting enriched U-235 93.2\% cylinders of
radius=10.15cm and height = 41.4cm in a aluminum containers of
thickness=0.15cm. The density of Aluminum is 2.71g/cc and density
of UO$_2$ F$_2$ water solution is 1.131g/cc. The cylinders were
set on equilateral triangle with 0.38cm surface separation. The
atomic density of material was 0.0021345 (U-235), 0.00015382
(U-238), 0.33383 (Oxygen), 0.65930 (Hydrogen) and 0.0045756
(Fluorine) in atoms/barn. The geometry for problem 11 is shown in
fig.\ref{godiva3}

k$_{eff}$(MONC)= 0.9993 $\pm$ 0.0003

k$_{eff}$(MCNP)=0.9991 $\pm$ 0.0011

\begin{figure}
\includegraphics[scale=0.45]{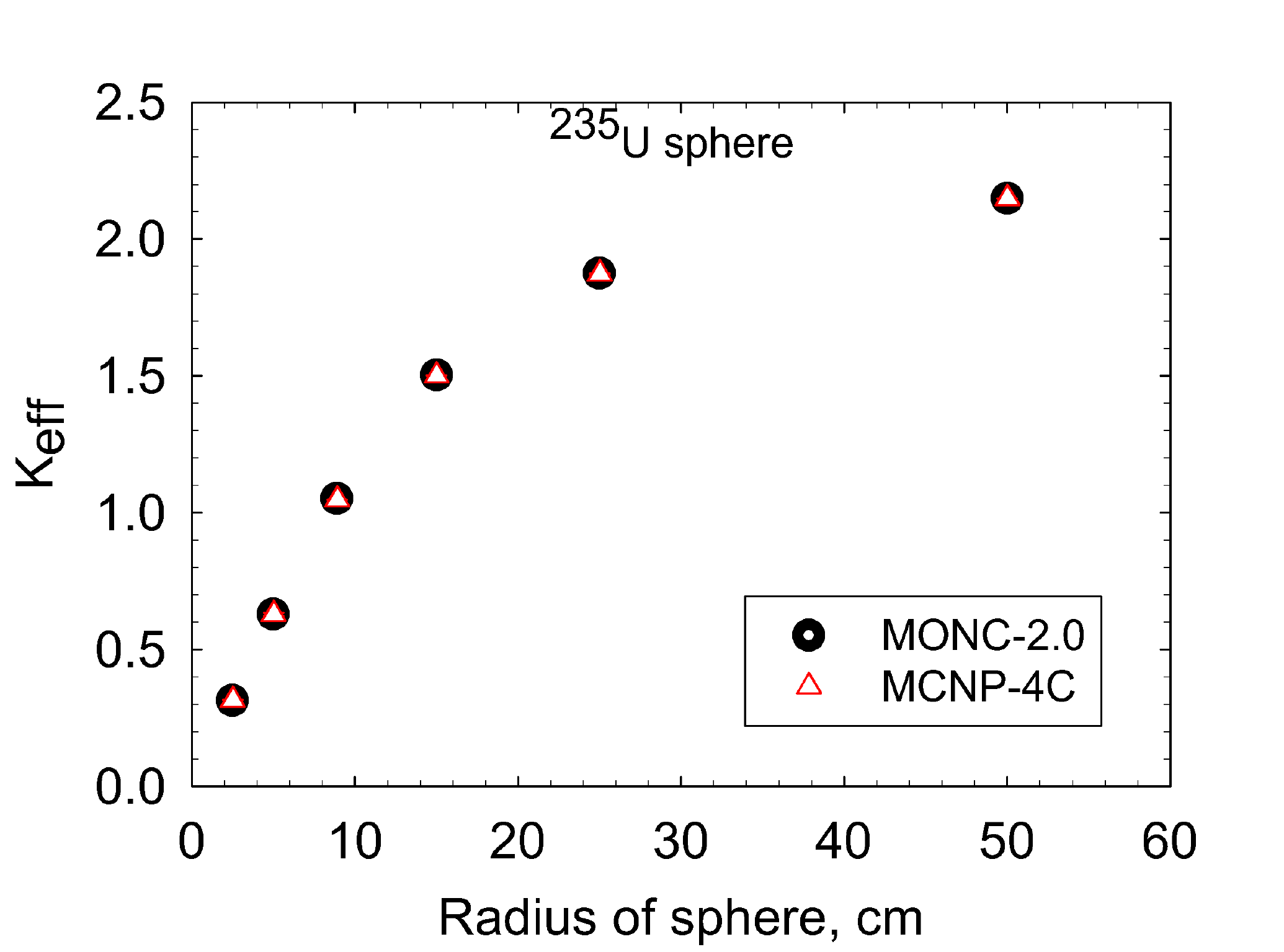}
\caption{k$_{eff}$ values for Sub-critical, Critical and
Super-critical $^{235}$U spherical assemblies} \label{godiva4}
\end{figure}

12)The $^{235}$U sphere of different raddi (2.5cm - 50cm) are
simulated for the sub-critical, critical and super-critical cases
to verify the algorithm of weight adjustment. The k$_{eff}$ values
are plotted in the Fig. \ref{godiva4}. Numerical values are given
in Table \ref{tab:keffur}.
\begin{table}
\caption{\label{tab:keffur} k$_{eff}$ values for different radii
of the $^{235}$U sphere. A comparison is performed with MONC-2.0
and MCNP-4c}
\begin{ruledtabular}
\begin{tabular}{cccc}
R,cm&k$_{eff}$ (MONC-2.0)&k$_{eff}$ (MCNP-4c)\\\hline
2.5& 0.3163$\pm$0.0014 & 0.3152$\pm$0.0003 \\
5.0& 0.6295$\pm$0.0057 & 0.6291$\pm$0.0001 \\
8.9& 1.0531$\pm$0.0002 & 1.0522$\pm$0.0001 \\
15.0& 1.5017$\pm$0.0004 & 1.5025$\pm$0.0006 \\
25.0& 1.8736$\pm$0.0003 & 1.8746$\pm$0.0004 \\
50.0& 2.1485$\pm$0.0002 & 2.1485$\pm$0.0003 \\
\end{tabular}
\end{ruledtabular}
\end{table}

\begin{figure}
\includegraphics[scale=0.45]{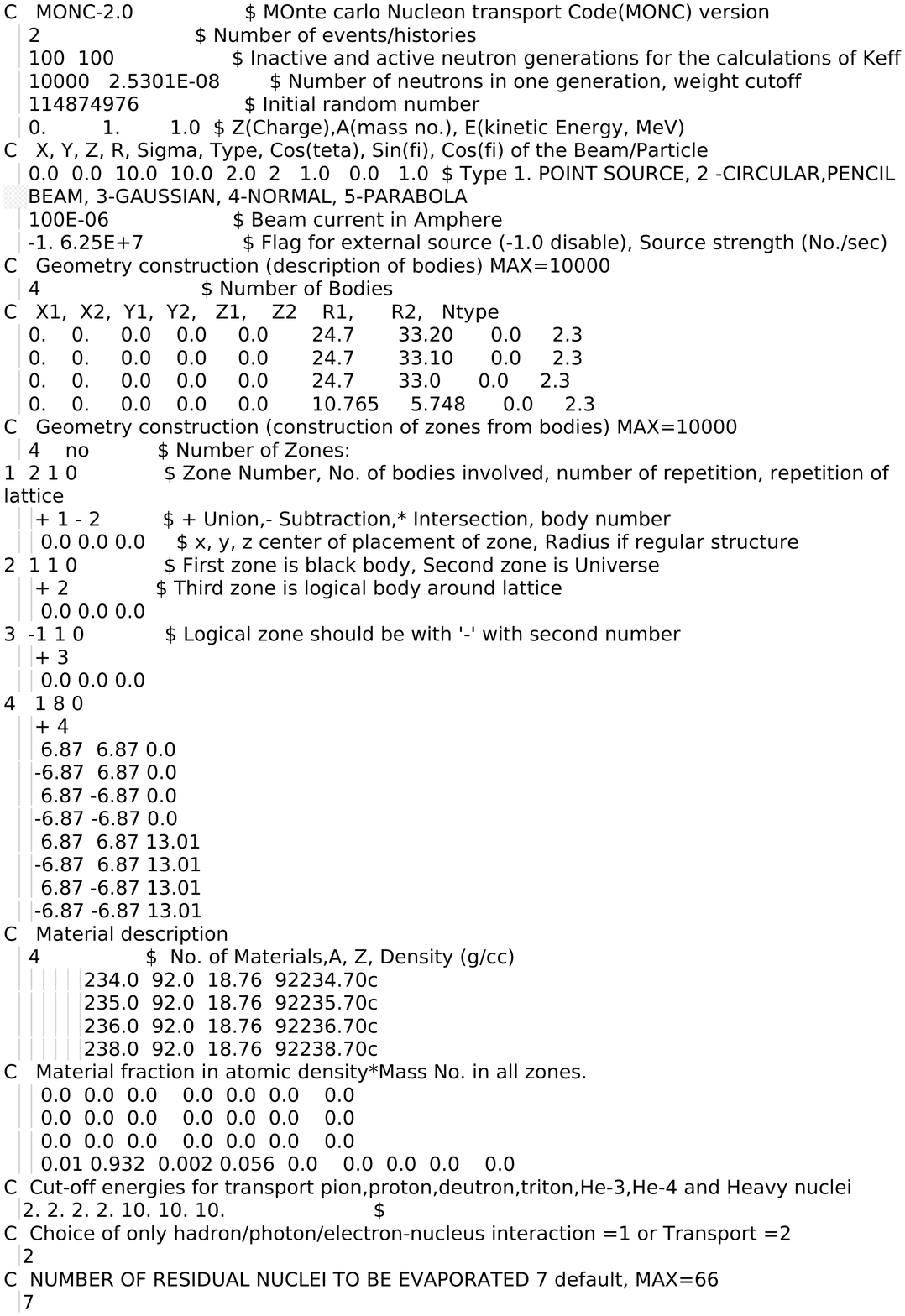}
\caption{MONC input file for 2x2x2 array of 93.2$\%$ enrich uranium cylinderical assembly.} \label{c8bareinput}
\end{figure}

\begin{figure}
\includegraphics[scale=0.45]{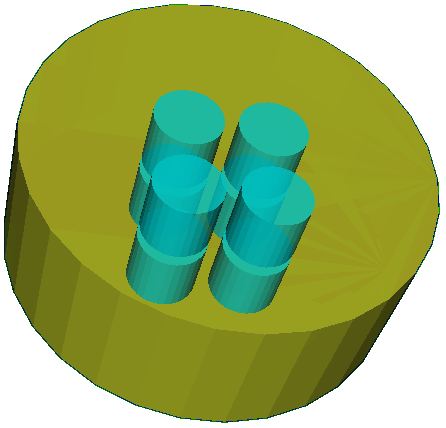}
\caption{2x2x2 array of 93.2$\%$ enrich uranium cylinderical assembly.} \label{c8bare}
\end{figure}

Problem $\#$13 A 2x2x2 un-reflected array of 93.2$\%$ enriched $^{235}$U metal cylinders as described in \cite {c8problem1} and shown in Fig. \ref{c8bare} is considered for k$_{eff}$ calculation. The surface separation in x-y direction is 2.244cm and it's 2.245cm in z-direction. Height and diameter of the cylinders are 10.765cm and 11.496cm respectively. The repeated structure of MONC is employed. The sample input file is given in Fig. \ref{c8bareinput}. 
k$_{eff}$(MONC)= 1.0012 $\pm$ 0.00121

k$_{eff}$(MCNP)=0.9999 $\pm$ 0.0009

k$_{eff}$(KENO)=0.9996 $\pm$ 0.0011

\begin{figure}
\includegraphics[scale=0.45]{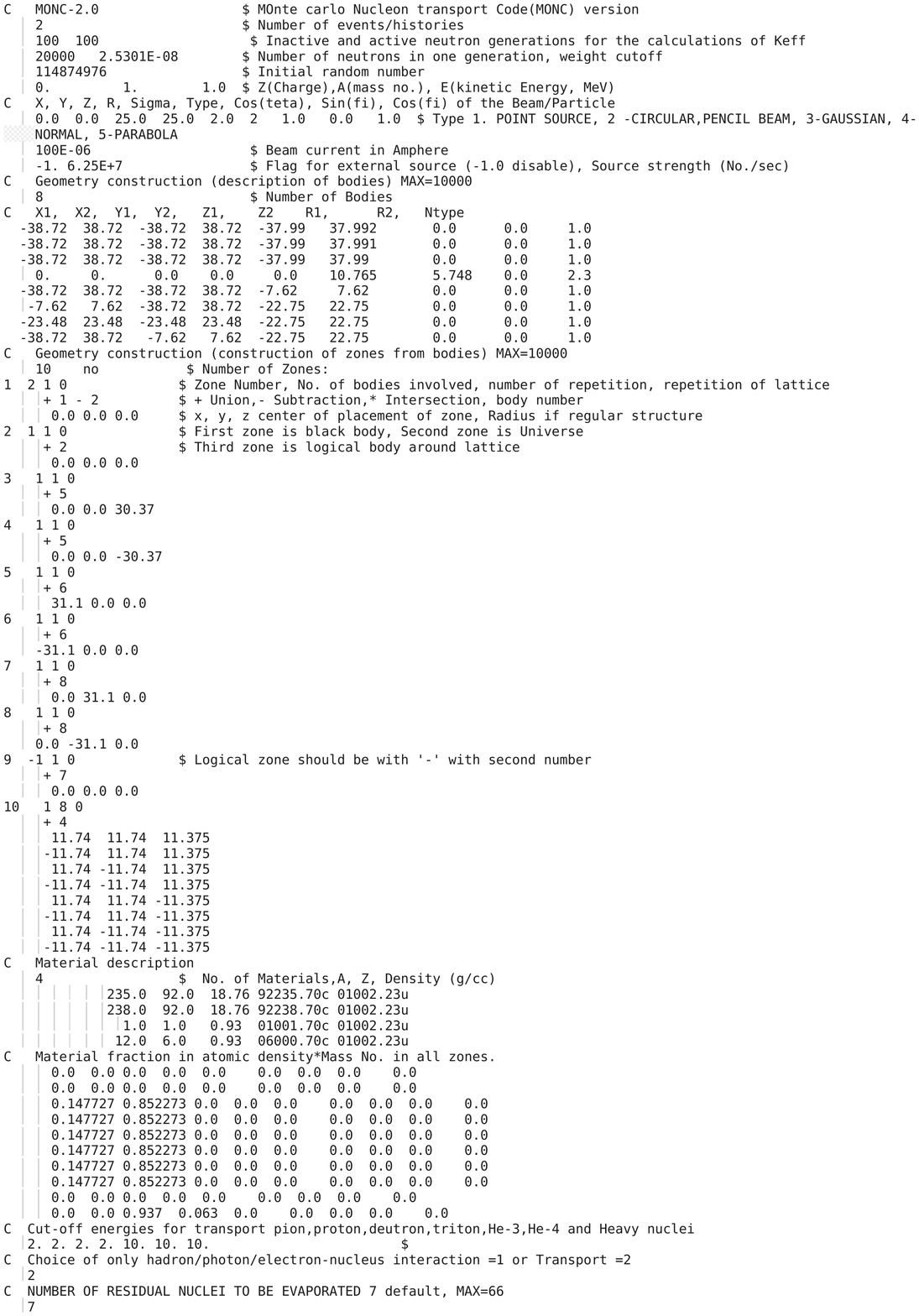}
\caption{MONC input file for 2x2x2 array of 93.2$\%$ enrich uranium cylinderical assembly, reflected by 15.24cm thick paraffin all around.} \label{c8paraffininput}
\end{figure}

\begin{figure}
\includegraphics[scale=0.35]{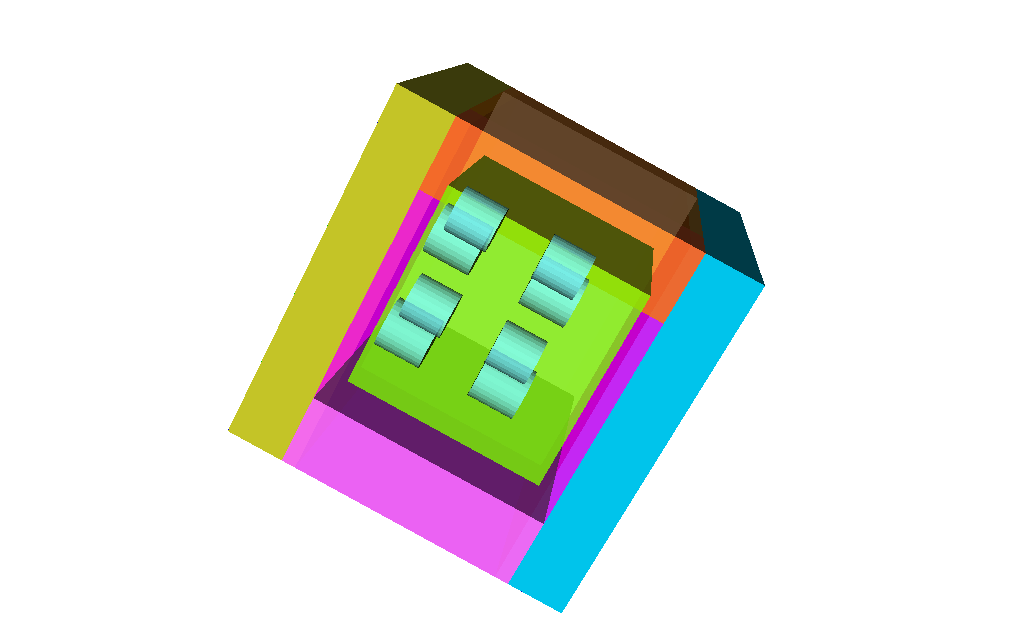}
\caption{2x2x2 array of 93.2$\%$ enrich uranium cylinderical assembly, reflected by 15.24cm thick paraffin all around.} \label{c8paraffin}
\end{figure}

Problem $\#$14 A 2x2x2 paraffin reflected array of 93.2$\%$ enriched $^{235}$U metal cylinders as shown in Fig. \ref{c8paraffin} is considered for k$_{eff}$ calculation. The surface separation in x-y direction is 11.984cm and it's 11.985cm in z-direction. Height and diameter of the cylinders are 10.765cm and 11.496cm respectively. The arrangement of uranium cylinders is surrounded by 15.24cm thick paraffin of density 0.93cm. The repeated structure of MONC is employed. The sample input file is given in Fig. \ref{c8paraffininput}. 
k$_{eff}$(MONC)= 0.99752 $\pm$ 0.00337

k$_{eff}$(MCNP)=0.9990 $\pm$ 0.0011

k$_{eff}$(KENO)=1.0009 $\pm$ 0.0013

Problem $\#$15 This is same problem as roblem $\#$14 except the paraffin thickness of 30.48cm instead of 15.24cm.  
k$_{eff}$(MONC)= 0.99752 $\pm$ 0.00337

k$_{eff}$(MCNP)=0.9995 $\pm$ 0.0027

k$_{eff}$(KENO)=1.0210 $\pm$ 0.0009

\subsection{Pincell model of PWR assembly}
\begin{figure}
\includegraphics[scale=0.6]{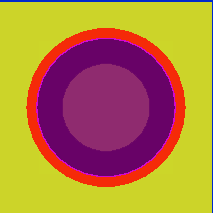}
\caption{pin cell model with reflective boundaries} \label{pincell}
\end{figure}

Single pincell model of PWR assembly was used to test MONC. Boundaries of the cell are assumed to be 
reflecting. The front view of the pincell model is shown in Fig. \ref{pincell}. The hight was taken as 4cm with reflecting from top and bottom. ENDF data libraries with 300K temperature was used for the simulation. The fuel was uranium oxide (UO$_2$) with 9.75\% enrichment. Pincel model parameters and initial fuel composition are given in Tables \ref{tab:pincellpar} and \ref{tab:pincellfuel}, respectively. The K$_{eff}$ value at t=0 is 1.568.
\begin{table}
\caption{\label{tab:pincellpar} Pin cell model parameters}
\begin{ruledtabular}
\begin{tabular}{cccc}
parameters        & Values    \\\hline
Fuel pellet radius (cm) &  0.4096 \\
Cladding inner radius (cm) & 0.4178 \\
Cladding outer radius (cm)  &  0.4750 \\
Pin pitch (cm)              &  1.26\\
Fuel density (g/cm 3 )      & 10.3\\
Cladding density (g/cm 3 )  &  6.550\\
Coolant density (g/cm )     &  0.997\\
Power density (kW/liter core) & 104.5\\
Specific power (W/gU)       &  34.6679\\
\end{tabular}
\end{ruledtabular}
\end{table}

\begin{table}
\caption{\label{tab:pincellfuel} Initial material composition}
\begin{ruledtabular}
\begin{tabular}{cccccc}
& Nuclide        & Weight (\%) & No. Density /cm$^3$    \\\hline
Fuel       & $^{234}$U      & 0.0688 & 1.82239E+19 \\
    &    $^{235}$U      & 8.5946 & 2.26826E+21 \\
     &   $^{238}$U      & 79.4866 & 2.07128E+22 \\
      &  $^{16}$O      & 11.8500 & 4.59686E+22 \\\hline
 Cladding      & O      & 0.125 & 3.08281E+20 \\
        & Cr      & 0.1 & 7.58663E+19 \\
        & Fe      & 0.21 & 1.48338E+20 \\
        & Zr      & 98.115 & 4.24275E+22 \\
        & Sn      & 1.45 & 4.81835E+20 \\\hline
Coolant        & $^{1}$H      & 11.19 & 6.66295E+22 \\
        & $^{16}$O      & 88.81 & 3.33339E+22 \\
\end{tabular}
\end{ruledtabular}
\end{table}


\subsection{Unit cell calculation for 19 and 37 fuel element in PHWR assembly}
\begin{figure}
\includegraphics[scale=0.6]{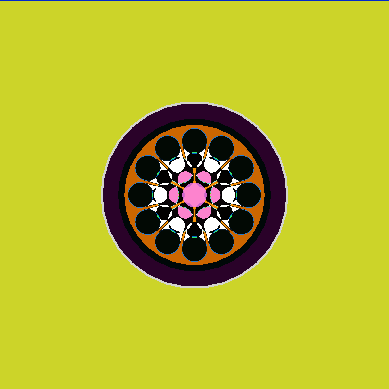}
\caption{PHWR-19 pin assembly} \label{phwr19}
\end{figure}

\begin{figure}
\mbox{\subfigure{\includegraphics[scale=0.25]{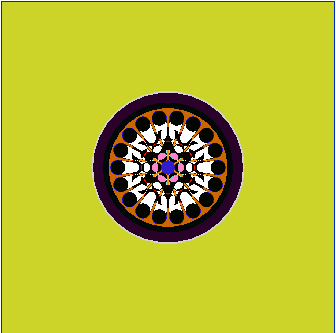}}}
\quad\subfigure{\includegraphics[scale=0.25]{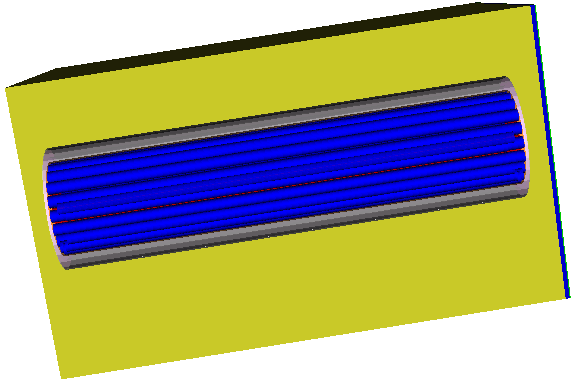}}
\caption{PHWR-37 pin assembly, front and side view} \label{phwr37}
\end{figure}

The details of the PHWR 19 and 37 rod fuel assembly are taken from Ref.\cite{baltej}
and are given below. The 19-rod fuel assembly has 1+6+12 central, first ring and second ring arrangement as shown in Fig. \ref{phwr19} and 37-rod fuel assembly has (1+6+12+18) structure as given in Fig. \ref{phwr37}. The dimensional and natural uranium fuel parameters for 19-rod fuel lattice assembly are given in Tables \ref{tab:phwr19cellpar} and \ref{tab:phwr19cellfuel}. The K$_{eff}$ values at t=0 without any load is 1.136 and 1.133 for 19 and 37 pin assemblies, respectively. Detailed burnup calculations are underway.

\begin{table}
\caption{\label{tab:phwr19cellpar} 19-rod unit cell model parameters}
\begin{ruledtabular}
\begin{tabular}{cccc}
parameters        & Values    \\\hline
Diameter of fuel rod (cm)              &  1.437\\
Fuel pellet (UO$_2$ ) diameter (including gap) & 1.445\\
Fuel element diameter (including sheath) &  1.521 \\
1$^{st}$ ring Pin Circle Diameter (cm) &  3.302 \\
2$^{nd}$ ring Pin Circle Diameter (cm) & 6.358 \\
Stack length (cm)  &  48.118\\
Fuel bundle length (inclusive both end plates)(cm )     &  49.53\\
Inner diameter of pressure tube (cm) & 8.26\\
Outer diameter of pressure tube (cm)       &  9.0\\
Air gap thickness (cm) & 0.85\\
Inner diameter of calandria tube (cm) & 10.80\\
Outer diameter of calandria tube (cm) & 11.06\\
Lattice spacing (square)(cm) & 22.86\\
\end{tabular}
\end{ruledtabular}
\end{table}
\begin{table}
\caption{\label{tab:phwr19cellfuel} 19-rod initial material composition}
\begin{ruledtabular}
\begin{tabular}{cccccc}
Material        & Density g/cm$^3$    \\\hline
Nat. UO$_2$       & 10.6 \\
Zr-4    &  8.91 \\
 pressure tube Zr-(2.5\% Nb)       & 8.91 \\
 Coolant D$_2$O  (at 271$^o$C)     & 0.8426 \\
 Calandria tube material Zr-2       & 8.91 \\
 Moderator D$_2$O (at 54.4 C)      & 1.0935 \\
 Specific power (KW/ kg of U) & 25.81 \\
\end{tabular}
\end{ruledtabular}
\end{table}

\begin{table}
\caption{\label{tab:phwr19cellfuel} 37-rod unit cell fuel element parameters}
\begin{ruledtabular}
\begin{tabular}{cccc}
parameters        & Values    \\\hline
Diameter of fuel rod (cm)              &  1.218\\
Fuel pellet-clad air gap (cm) & 0.008\\
Fuel pellet (UO$_2$ ) diameter (including gap) & 1.308\\
Fuel element diameter (including sheath) &  1.521 \\
1$^{st}$ ring Pin Circle Diameter (cm) &  2.978 \\
2$^{nd}$ ring Pin Circle Diameter (cm) & 5.750 \\
3$^{rd}$ ring Pin Circle Diameter (cm) & 8.660 \\
Stack length (cm)  &  48.118\\
Fuel bundle length (inclusive both end plates)(cm )     &  49.5\\
Inner diameter of pressure tube (cm) & 10.338\\
Outer diameter of pressure tube (cm)       &  11.238\\
Air gap thickness (cm) & 1.694\\
Inner diameter of calandria tube (cm) & 12.932\\
Outer diameter of calandria tube (cm) & 13.212\\
Lattice spacing (square)(cm) & 28.60\\
Specific power (kW/ kg of U) & 26.58 \\
\end{tabular}
\end{ruledtabular}
\end{table}

\section{\label{sec:level7} Decay of the Radio-active element}
Monitoring of decay heat, radio-activity of spent fuel, fission products and minor Actinides is an important task for spent fuel management, reprocessing and permanent disposal.The decay and burnup calculations are required for estimation of inventory of accelerator or reactor
produced nuclear waste. The generalized Bateman's eq. \cite{bateman06} is solved exactly with linear chain method. This kind of method is also used in other codes like CINDER \cite{cinder98} and BISON \cite{bison} while ORIGEN \cite{origin} uses matrix exponential method to solve the equations. The Bateman equation for decay in n-nuclide series in linear chain describing n$^th$ nuclide concentration at time t is given by
\begin{equation}
 N_{n}(t)=\frac{N_1(0)}{\lambda_n}\sum_{i}^n \lambda_{i}\alpha_{i}exp(-\lambda_{i}t) \label{bateman1}
\end{equation}
here
\begin{equation}
 \alpha_i=\prod_{\substack{j=1\\j\neq i}}^n \frac{\lambda_j}{(\lambda_j-\lambda_i} \label{bateman2}
\end{equation}

Here N$_1$(0) $\neq$ 0 and  N$_i$(0) = 0 when i $>$ 1, means inventory of all daughters are assumed zero at time zero, $\lambda_{i}$ is the decay constant of i$^{th}$ nuclide. This equation is valid if the decay constants are different otherwise it becomes infinite. Very small difference in decay constants also may lead to biased numerical results. We have artificially shifted the decay constants to get the approximate solutions as these kind of problems are not with too many isotopes in the chain. 
In case of transmutation due to particle flux, the modified decay constant is used as follows
\begin{equation}
    \lambda_{i,j}=b_{i,j}.\lambda_j + \sum _{x=n,p,\pi,..} \int\phi^x\sigma_{i,j}^x(E)dE \label{bateman3}
\end{equation}
where b$_{i,j}$ is is branching ratio of decay through some decay channel of nuclide j to nuclide i. $\phi^x$ is particle flux and $\sigma_{i,j}^x$ is production cross-section of nuclide i through reaction with j by particle x. In other words one can replace the decay constant $\lambda \Longrightarrow \lambda$ + $\sigma\phi$. 
ENDF-B decay data library is used to extract decay properties (half life, decay type ($\beta^-$, $\beta^+$, $\alpha$, EC, decay heat, decay spectra etc.) of the radio-active/stable isotopes. 

\begin{figure}
\includegraphics[scale=0.4]{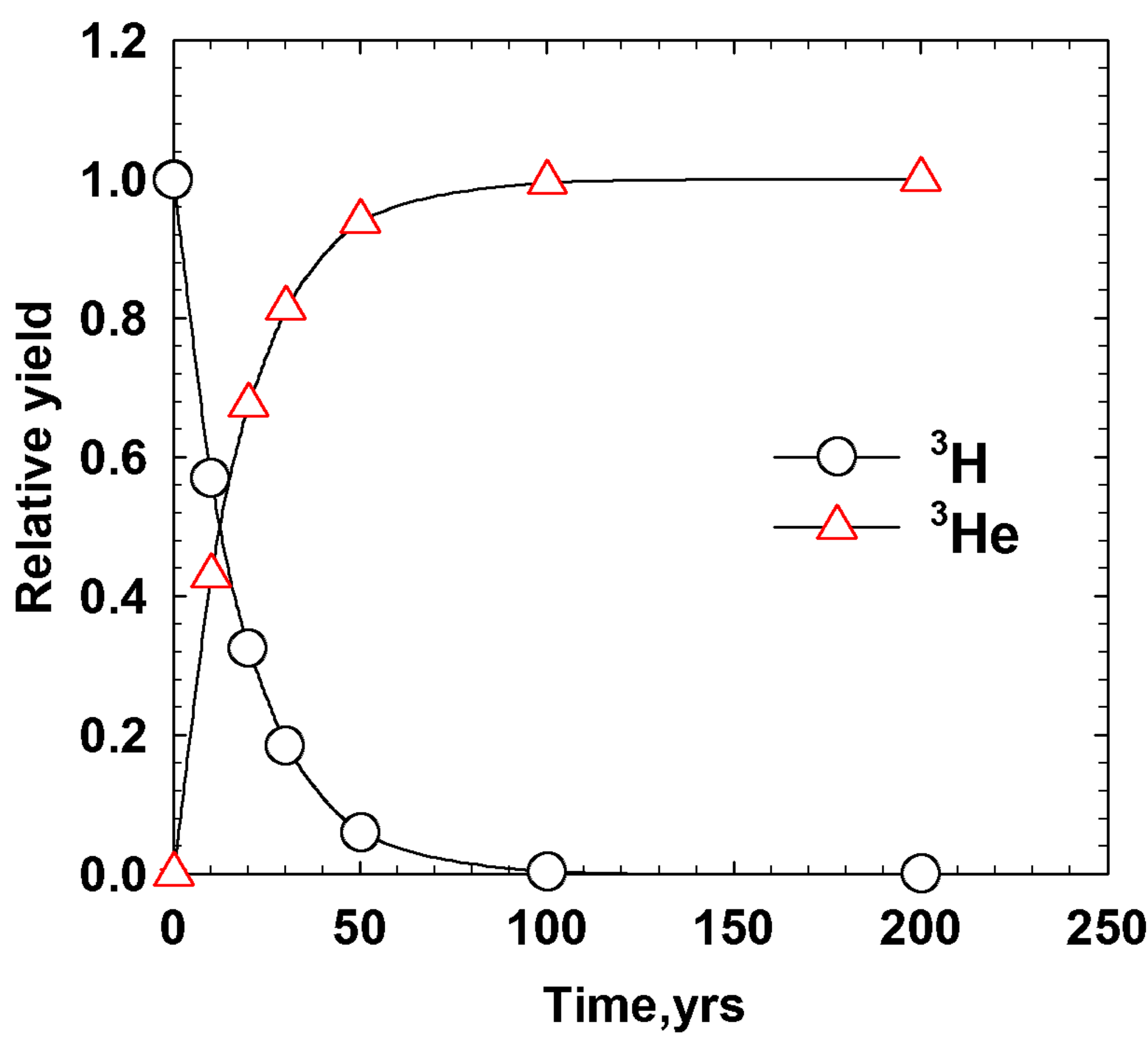}
\caption{Decay of $^3$H} \label{decayh3}
\end{figure}

\begin{figure}
\includegraphics[scale=0.4]{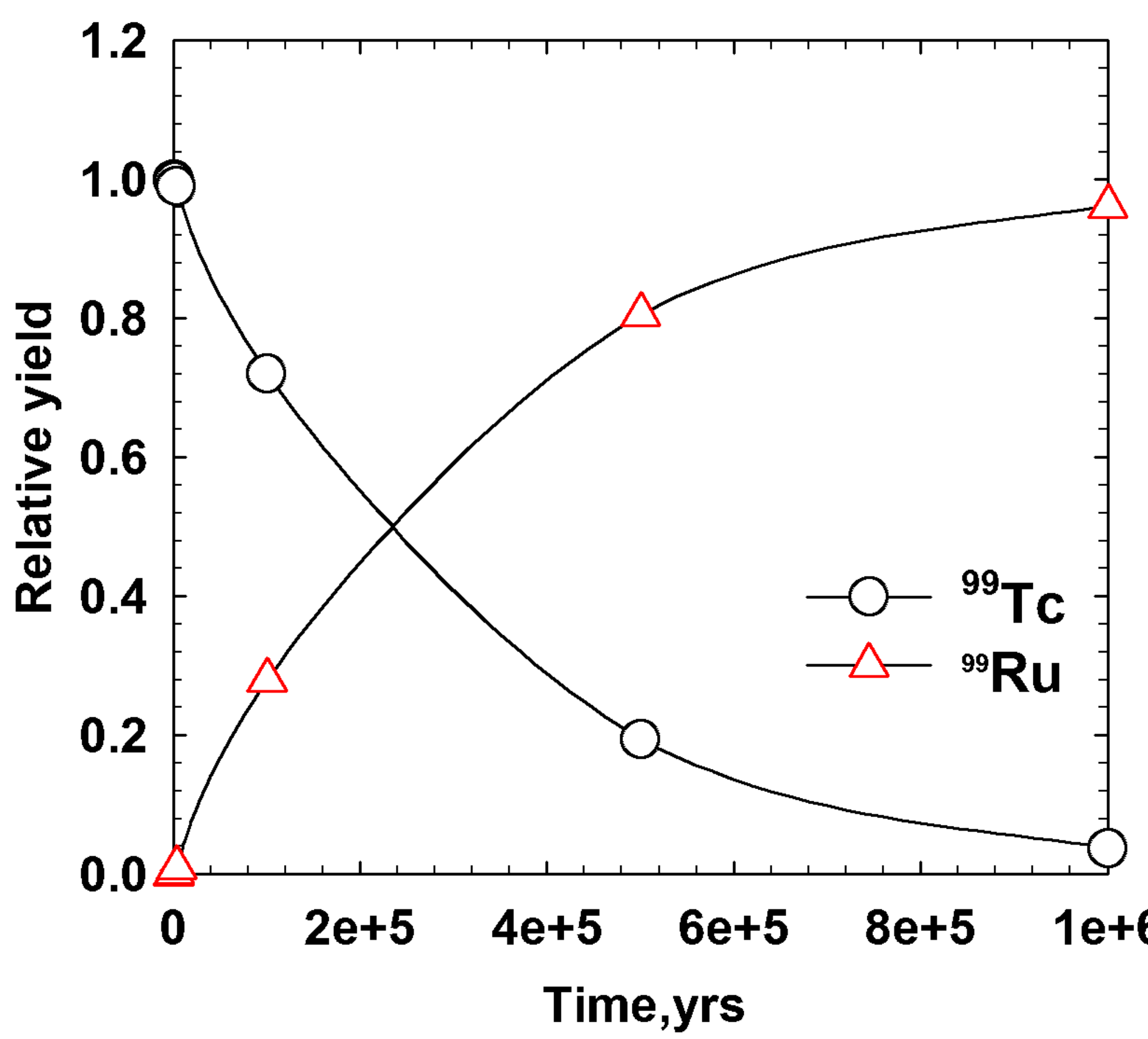}
\caption{Decay of $^{14}$C} \label{decayc14}
\end{figure}

\begin{figure}
\includegraphics[scale=0.4]{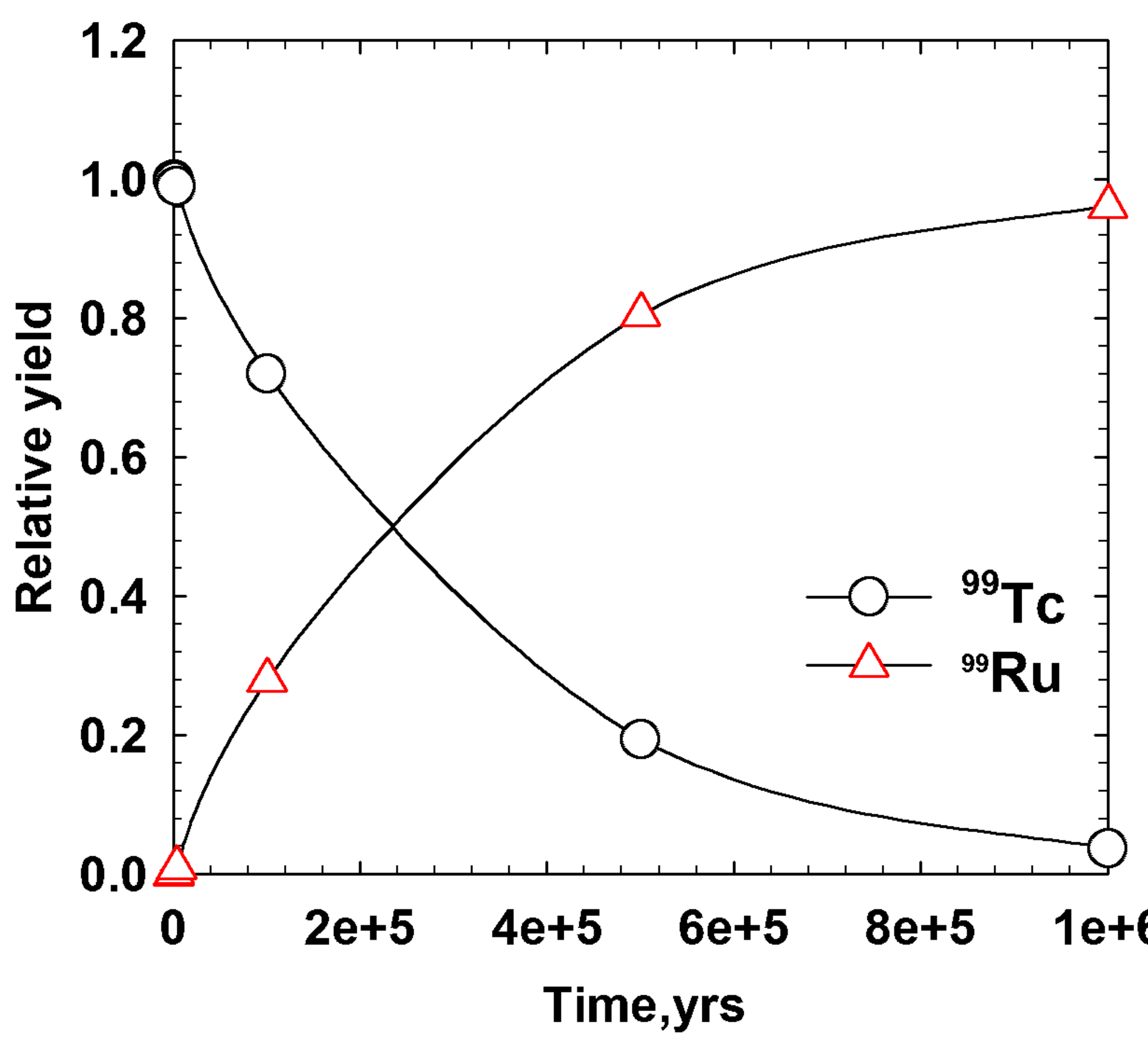}
\caption{Decay of $^{99}$Tc} \label{decaytc}
\end{figure}

\begin{figure}
\includegraphics[scale=0.4]{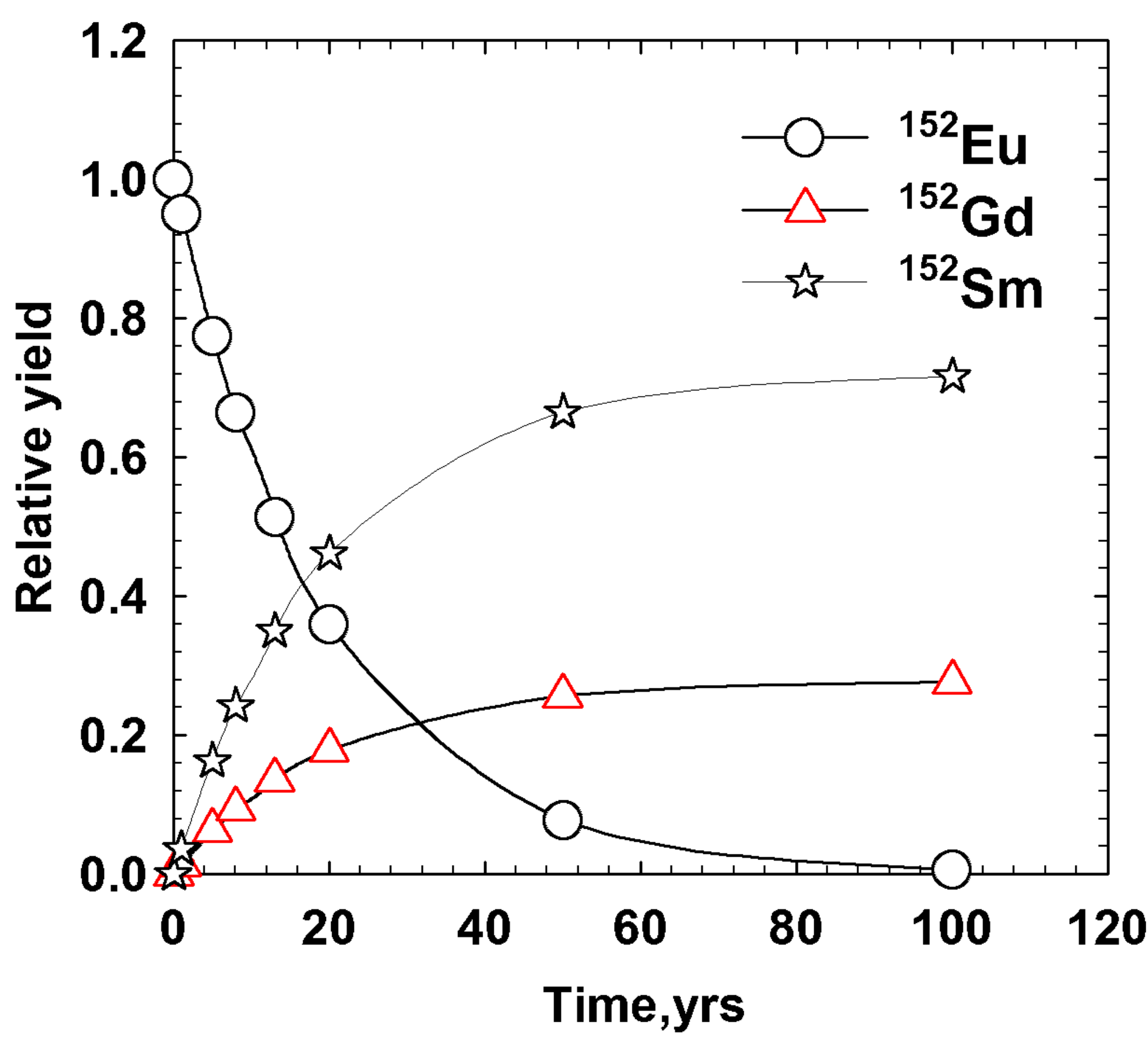}
\caption{Decay of $^{152}$Eu} \label{decayeu152}
\end{figure}

\begin{figure}
\includegraphics[scale=0.4]{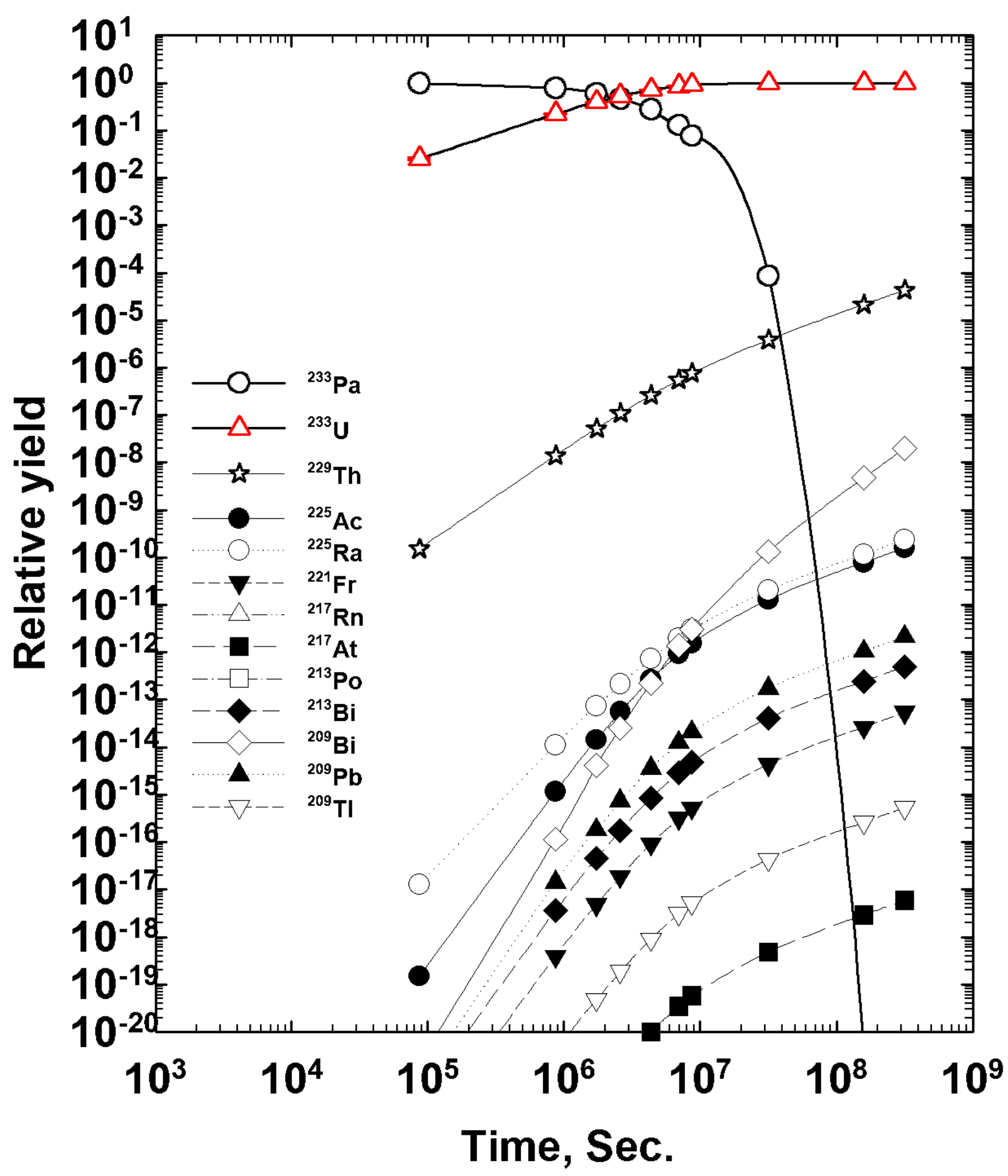}
\caption{Decay of $^{233}$Pa} \label{decaypa233}
\end{figure}

Deacy of $^3$H,$^{14}$C,$^{99}$Tc,$^{152}$Eu $^{233}$Pa are given in Fig. \ref{decayh3},\ref{decayc14},\ref{decaytc}, \ref{decayeu152} \ref{decaypa233}. 

One group cross-sections are calculated for transmutation/burnup calculations. (n,$\gamma$), (n,f), (n,2n), (n,3n), (n,4n), (n,$\alpha$), (n,p) reactions are included in the burnup. Flux normalization is used and normalization factor is  based on the following equation \ref{fmf}.

\begin{equation}
 FN=\frac{P}{\sum_{i}^{n}V_i\sum_j^{m_i} N_{ij}Q_j\int \sigma_{ij,f}(E)\phi_i(E)dE} \label{fmf}
\end{equation}
where P is total power,\\ 
where V$_i$ is volume of zone i,\\ 
N$_{ij}$ is number density of actinide j in zone i,\\ 
Q$_{j}$ is recoverable energy of actinide j,
$\sigma_{ij,f}$(E) is fission cross-section for actinide j in zone i.\\

or alternatively by the following eq. \ref{fmf2}.
where Q(Z,A)(MeV/fission) = 1.29927E-3 x Z$^2$ A$^{0.5}$ + 33.12

\begin{equation}
 FN=\frac{P.\nu}{Q.K_{eff}} \label{fmf2}
\end{equation}

\section{\label{sec:level6} Graphical User Interface}
\begin{figure}
\includegraphics[scale=0.16]{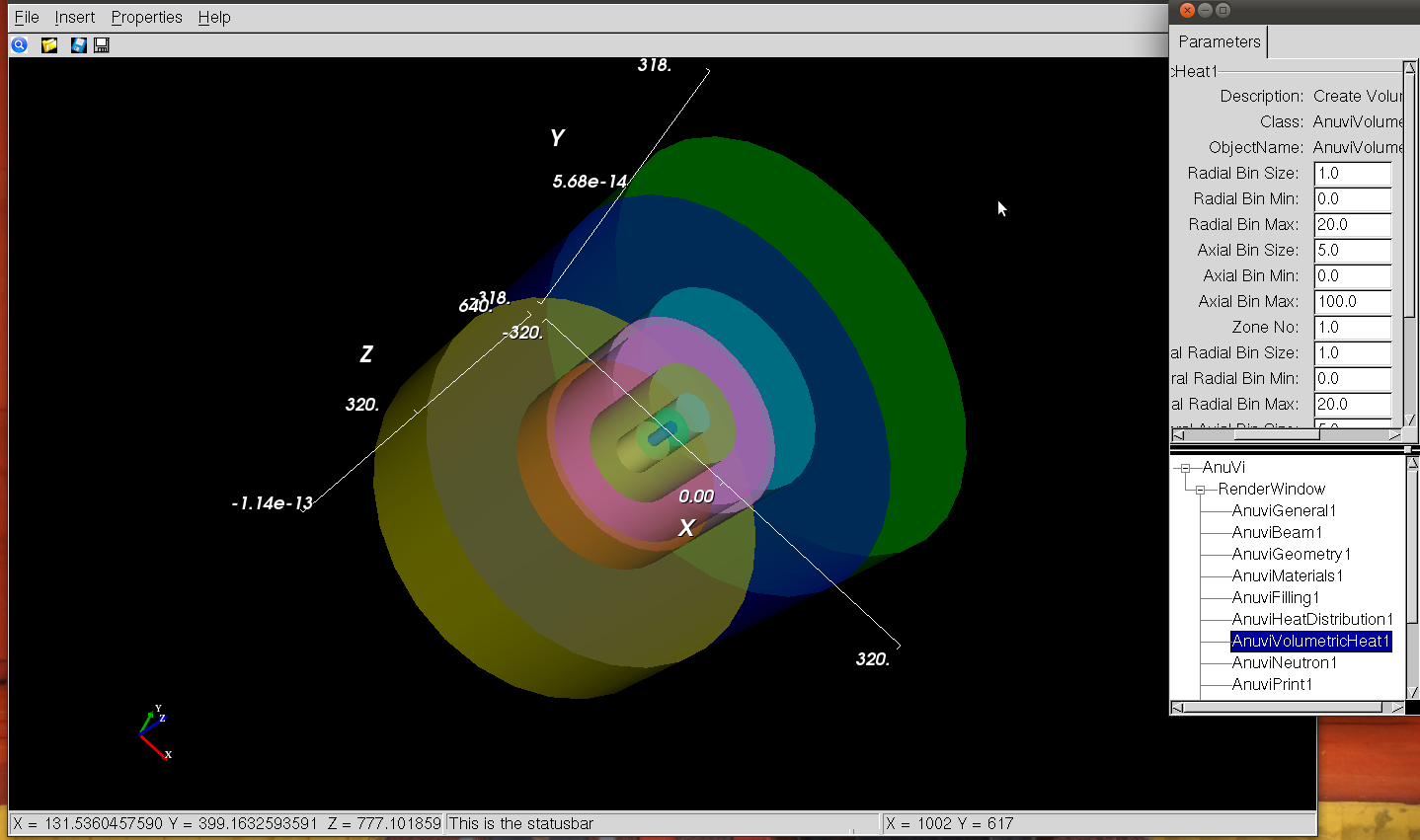}
\caption{GUI showing concentric geometrical assembly} \label{gui1}
\end{figure}

Graphical User Interface (GUI) along with data visualization is a powerful tool 
required for supporting such ambitious software. The GUI and the visualization
modules are developed to support cross-platform usage. Development of these 
modules is done in Python language using the base libraries of Visualization 
Toolkit [9] for visualization and WxPython for GUI. The communication between 
the GUI and the Monte-Carlo code is through loose coupling, i.e. both these 
modules are independent of each other and the communication is through external files. 
The GUI and visualization modules are developed for cross-platform usage, 
so that they can be run on all Windows and Linux platforms.
Construction and display of geometry, making the input file (defining the 
material properties and filling the material in different zones, defining the number of 
events and description of the output tallies etc.) and analysis of the output 
data are handled in this GUI. A snapshot of the MONC GUI depicting multiple 
concentric cylinders is shown in Fig.\ref{gui1} running under 32-bit Ubuntu Linux environment.

To construct geometry, all Boolean operations viz. union, subtraction, intersection 
are available in this framework to make the complex zones from the basic bodies. 
Scaling, rotation and translation of the basic bodies is supported here. 
This information is saved in the input file and the Monte-Carlo code can be run 
either through terminal or from the GUI itself. Standard features viz. showing 
3D-axis around bodies, taking snapshots are provided.
Boolean operations Union, subtraction and intersection are mapped to operators +,- 
and * respectively, the construction of a geometry is mapped to a simple mathematical expression. 
Computational Geometry Algorithm Library (CGAL) is used for implementing 
the Boolean operations and converting them to equivalent triangulated meshes. 
The CGAL created meshes are converted to equivalent VTK filter objects and are 
joined into the VTK rendering pipeline consisting Mapper, Actor, Renderer and RenderWindow objects. 
Currently the code supports only five basic building bodies namely box, Cone, 
cylinder, hexagon and sphere. From these five basic building bodies, any number 
of complex geometries using Boolean operations. For visualization, tessellation 
of the geometry is mandatory and thus the basic building blocks are all complex 
polyhedrons in nature. A complex polyhedron is defined as a closed object whose 
boundary is manifold i.e. in the Euler Characteristic equation for polyhedrons 
X=V-E+F, X must be equal to 2, where V is the number of vertices forming polyhedral, 
E is the number of edges and F is the number of faces forming the polyhedral. 
Though Python being an interpreter based language is slower compared to performance 
of the compiled executable files, it is used owing to its inherent features supporting 
rapid application development. In-order to offset the inferior performance of the 
Python running mathematically intense calculations, the classes which are intense 
in terms of computations are developed in C++ and are compiled into dynamic libraries. 
Using Python-C++ wrapper libraries information is exchanged between C++ and Python and vice-versa.
Cross-platform development is done in two phases. 

The first phase, i.e. 
the development of code is done using the ANSI standards of the C++, 
so that the compilation is smooth on all platforms like Windows and Linux. 
The second phase is to support compilation procedure on all platforms. 
The Linux Make files and Windows Project files have to be created properly. 
This is achieved by writing higher level CMake scripts. 
The Menu structure is divided into four groups namely File, Properties, 
Insert and Help. File menu consists of opening and saving of the MONC scripts 
as well as saving the screen shots. Properties menu handles all the property 
pages which can be modified by the users for supporting various simulations. 
Insert menu gives control over the insertion of bodies and construction of 
geometries using the complex mathematical expressions consisting of multiple
Boolean operations. Help menu provides a detailed help system which is developed 
using the advanced HTML help engine supporting searching of the key words, indexing 
of the help items as well. A tool bar useful with frequently used operations is provided. 
Standard three button mouse interactions are implemented viz. left-button controls the 
rotation, middle-button is for translation and the right-button is for scaling.
The most important aspect of GUI is to visualize and correct the geometry in 3-dimensions. 
Overlapping regions (if any) after scaling, rotation, and translation can be easily identified and corrected before running the Monte-Carlo code.

\section{\label{sec:level8} Conclusion}
The Monte Carlo code MONC has been developed for ADS, Spallation reactions, reactor, 
dosimetery, and shielding applications. New CSG model with Union, Subtraction and 
Intersection Boolean operations is developed to make the heterogeneous zones. Scaling, 
rotation, and translation operations are used to make more complex zones. 
Repeated geometry model has been developed for simulation of any complex reactor designs 
as well as the detector simulations. The pointwise cross section data for neutrons below 
20MeV are used. The S($\alpha$, $\beta$) scattering matrices for neutron 
energy < 4eV is used if it is available in the library for the given compound element 
otherwise Fermi-gas treatment is used. The code has been benchmarked for K$_{eff}$ of many 
experimental simple assemblies and is under extensive benchmark for different other 
assemblies including real Thorium Plutonium MOX fuel based AHWR system. List of 
fission product and their spatial distribution can be analyzed using this code at any given time. MONC uses MPI architecture and has been tested for more than 1000 nodes.
\end{document}